\documentclass[journal]{IEEEtran}
\usepackage{amsmath,amsfonts,amssymb}

\usepackage{algorithm}
\usepackage{algpseudocode}

\usepackage{booktabs}
\usepackage{multirow}
\usepackage{tabularx}
\usepackage{diagbox}

\usepackage[nocompress]{cite}
\usepackage{hyperref}

\usepackage[caption=false,font=footnotesize]{subfig}
\usepackage{graphicx}

\usepackage{tikz, pgfplots}
\usetikzlibrary{shapes, arrows, arrows.meta, fit, positioning, calc, 3d}
\usepgfplotslibrary{fillbetween}
\pgfplotsset{compat=1.18} 
\tikzset{>=latex}

\newtheorem{theorem}{Theorem}
\newtheorem{lemma}{Lemma}
\newtheorem{remark}{Remark}
\newtheorem{definition}{Definition}
\newtheorem{assumption}{Assumption}
\newtheorem{proposition}{Proposition}

\allowdisplaybreaks

\begin{document}
\bstctlcite{IEEEtran:no_dash_repeated_names}

\title{Value of Communication in Goal-Oriented Semantic Communications: A Pareto Analysis}
\author{Jiping~Luo,~Bowen~Li,~and~Nikolaos~Pappas,~\IEEEmembership{Senior~Member,~IEEE}
\thanks{The authors are with the Department of Computer and Information Science, Link\"oping University, 58183 Link\"oping, Sweden (e-mail: jiping.luo@liu.se; bowen.li@liu.se; nikolaos.pappas@liu.se). This work was supported by ELLIIT, the Graduate School in Computer Science (CUGS), and the European Union (6G-LEADER) under Grant 101192080.}
}

\maketitle
\begin{abstract}
Emerging cyber-physical systems increasingly operate under stringent communication constraints that preclude reliable transmission of all available machine-type data. Motivated by this challenge, goal-oriented semantic communication advocates a minimalist design principle: transmit only what is necessary to achieve the system's goal. In this work, we formulate optimal semantic communication design as a bi-objective Markov decision process (MDP) that trades off two competing objectives: system performance and communication cost. In contrast to classical approaches that seek to optimize system performance by exhausting a prescribed communication budget, we propose a minimalist design that answers: What is the marginal value of communication, and what is the minimum communication required to achieve the goal? Our approach is based on a Pareto analysis that characterizes the complete set of policies achieving optimal tradeoffs between these two objectives. The value of communication is defined as the absolute slope of the resulting Pareto front. A key result of this paper shows
that this front admits a tractable structure: it is convex and piecewise linear, and its corner points correspond to simple deterministic policies. The entire front can be constructed by mixing the deterministic policies at neighboring corner points. Leveraging these geometric properties, we introduce SPLIT, an efficient and provably optimal algorithm for computing the Pareto front. Numerical results demonstrate the efficiency of SPLIT, the diminishing returns of over-provisioning in communication, and the effectiveness of Pareto-optimal semantic communication design.
\end{abstract}

\section{Introduction}
Virtually all modern wireless systems are built upon fundamental principles of reliable communications first developed in Shannon's seminal work~\cite{shannon}. Under this classical view, ``reliability" primarily refers to the \emph{accuracy} of signal reconstruction. However, emerging cyber-physical systems increasingly operate under stringent communication constraints that preclude reliable transmission of all available machine-type data. Since raw measurements often contain correlated or redundant components, effective operation depends not on transmitting all available data but on selecting the information that contributes to the objectives of the system. Beyond accuracy, goal-oriented semantic communication assesses the \emph{value of information} and aims to generate and transmit only what is relevant and at the right time~\cite{howard1966information, kountouris2021semantics, luo2025cost}.

The value of information depends on the application context and goals. In causal systems, this value can be inferred from the history of all past system realizations~\cite{luo2025cost}, in contrast to distortion measures~\cite{gallager1968information}, which are typically history-independent. The notion of \emph{age} of information (AoI) is the first concrete step in this direction, built on the premise that information holds the greatest value when it is fresh~\cite{kaul2012real, kosta2017age}. Yet, this assumption deserves reexamination, as the freshest measurement may be inaccurate or uninformative. Recently, a series of semantics-aware metrics has been introduced that go beyond accuracy and age, capturing aspects such as context-awareness~\cite{stamatakis2019control, kountouris2021semantics, luo2025semantic}, the lasting impact of consecutive errors~\cite{maatouk2020age, salimnejad2023state, luo2024exploiting}, and the urgency of lasting impact~\cite{luo2025cost, luo2026role}. 

Despite this progress, the \emph{value of communication} has yet to be satisfactorily characterized. In the existing literature, one well-studied question is the following: given a fixed communication budget, how should it be used to optimize system performance? Analytical tools from Markov decision processes (MDPs), such as Lagrangian and constrained formulations, play a central role in addressing this problem (see~\cite{luo2025information} and the references therein). This approach, however, often communicates more information than necessary. Semantic communication inherently advocates a \emph{minimalist} design, yet a fundamental question remains: \emph{What is the marginal value of communication, and what is the minimum communication required to achieve the goal?} Answering this question requires a bi-objective formulation that jointly optimizes the two competing objectives, thereby providing a comprehensive view of their tradeoff.

\begin{figure}[t!]
    \centering
    \includegraphics[width=0.95\linewidth]{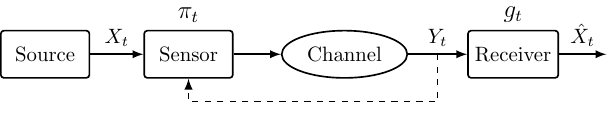}
    \caption{Remote state estimation of a Markov source.}
    \label{fig:system_model}
\end{figure}

In this work, we characterize the value of communication through remote state estimation of Markov sources, a fundamental problem in information and control theory. As depicted in Figure~\ref{fig:system_model}, a sensor observes a Markov source $X$ and decides when to send its current observation to a receiver over a wireless channel. The receiver employs the maximum a posteriori (MAP) estimation policy $g$ to reconstruct the source based on all received messages. The sensor seeks a communication policy $\pi$ to achieve an optimal tradeoff between the estimation performance and communication cost. Unlike the classical notion that estimation quality is synonymous with accuracy, we filter out less important measurements by quantifying their value through the urgency of lasting impact metric~\cite{luo2025cost}.

\subsection{Main Contributions}
In this work, we formulate optimal semantic communication design as a bi-objective MDP. Our main contributions are as follows.
\subsubsection{Pareto Formulation} In contrast to the Lagrangian and constrained formulations, which yield a single ``best" policy, the Pareto formulation seeks to characterize the complete set of optimal tradeoff policies (i.e., the Pareto front) that balance the conflicting estimation and communication costs. The value of communication is defined as the absolute slope of this front, quantifying the marginal improvement in estimation performance per unit increase in communication. 

\subsubsection{Exact Characterization of the Pareto Front} Computing the Pareto front for bi-objective MDPs is a nontrivial task. It requires solving infinitely many constrained optimization problems across all possible communication budgets. To the best of our knowledge, no practical algorithms currently exist to compute the complete Pareto front with formal guarantees. A common approach is to approximate the front through a dense search over the entire budget range. Nevertheless, this approach is inefficient and provides only a coarse characterization of the value of communication.

A key contribution of this paper is an exact characterization of the Pareto front for bi-objective MDPs. We show that the front is strictly decreasing, convex, and piecewise linear. Each corner point corresponds to a stationary deterministic policy, and every point on the front can be realized by mixing the policies at neighboring corner points. We derive a closed-form expression for the mixing coefficient. The value of communication is characterized by a set of Lagrange multipliers. The theoretical results are validated through simulations on the semantic communication model. Numerical results show that the marginal value of communication diminishes at relatively high communication frequencies and that the minimum communication required to achieve a target performance level can be significantly lower than what the constrained formulation would consume.

\subsubsection{Implications of the Main Results}
The randomized mixing policy provides a complete and efficient representation of the Pareto front. Once the deterministic policies at the corner points are computed, we can construct the entire front exactly by mixing policies at adjacent corner points. Notably, this mixing policy is easy to implement: the corner policies are precomputed and stored in a lookup table, and when the operating budget or performance goal changes, the sensor simply selects the corner policies and adjusts the mixing coefficient. Thus, there is no need to resolve the optimization problem or to store infinitely many policies. This feature is useful in resource-constrained systems such as sensor networks, where devices operate under stringent computation and memory limitations.

\subsubsection{Structure-Aware Algorithm} We leverage these structural results to develop SPLIT, an efficient and provably optimal algorithm. SPLIT computes the Pareto front using a number of MDP solves that is linear in the number of corner points. Numerical results show that SPLIT computes the exact Pareto front with the same order of complexity as solving a single constrained problem. In contrast, existing approximation methods require more than an order of magnitude more MDP solves to obtain a close approximation.

\subsection{Organization of the Paper} 
The rest of the paper is organized as follows. Section~\ref{sec:related work} reviews related work on semantic communications and multi-objective MDPs. Section~\ref{sec:model} describes the semantics-aware communication model. Section~\ref{sec:formulation} formulates the optimal communication design problem as a bi-objective MDP and presents the Pareto, constrained, and Lagrangian formulations. Section~\ref{sec:pareto} establishes the structural properties of the Pareto front for bi-objective MDPs and discusses the relationships among the three formulations and the practical implications of the results. Section~\ref{sec:computing} presents the SPLIT algorithm for computing the Pareto front. Numerical results and concluding remarks are provided in Sections~\ref{sec:simulations} and~\ref{sec:conclusion}, respectively.

\section{Literature Overview}\label{sec:related work}
\subsection{Goal-Oriented Semantic Communications}\label{sec:related work 1}
Information \emph{accuracy} has long been the primary performance metric in information and control theory~\cite{shannon, gallager1968information, schenato2007foundations}. The rate-distortion theorem, however, makes clear that many emerging cyber-physical systems cannot reliably communicate the massive volumes of machine-type data they generate. From a data significance perspective, they need not do so: raw measurements often contain substantial superfluous information that is correlated, redundant, or irrelevant to the system's goals.

\emph{Timeliness} has emerged as a crucial requirement in applications such as intelligent transportation, industrial automation, and swarm robotics~\cite{luo2023real, vitturi2013industrial, gielis2022critical}. The AoI was introduced in~\cite{kaul2012real} to quantify the freshness of the information that a remote receiver has about a source. Reviews of AoI and its applications can be found in~\cite{yates2021age, sun2022age}, and~\cite{pappas2023age}. It is worth noting that the impact of aged information on networked control systems (NCSs) has been independently studied from a control-theoretic perspective~\cite{sinopoli2004kalman, schenato2008optimal}. In particular, a Kalman filter produces estimates of a linear Gaussian process and pushes them into a data queue for transmission. For control stability, it is unnecessary to transmit every queued packet, since the newest estimate is as accurate as the older ones and best reflects the current state of the system; see, e.g.,~\cite{shi2011sensor, leong2017sensor, liu2022remote, luo2025remote}. This canonical example illustrates the basic idea of AoI: when the newest measurement best reflects the system status, outdated information can be discarded to ensure timely operation at the destination. However, this is not always the case. Recent studies have shown that age-based policies are often suboptimal for monitoring general information sources, such as Wiener processes~\cite{sun2020sampling}, Ornstein-Uhlenbeck processes~\cite{ornee2021sampling}, and Markov chains~\cite{stamatakis2019control, salimnejad2023state, luo2025semantic}.

Several AoI variants have been proposed to balance timeliness and accuracy. Effective age~\cite{kam2018towards} penalizes the cumulative error incurred in the absence of updates. Similar in spirit, age of incorrect information (AoII)~\cite{maatouk2020age} and cost of memory error~\cite{salimnejad2023state} measure the duration for which the system remains in erroneous states. Subsequent work has analyzed these metrics in systems with random channel delays~\cite{chen2024minimizing}, multiple access~\cite{munari2024monitoring}, and continuous-time Markov sources~\cite{cosandal2024modeling, cosandal2025multi}, among others. Other variants, such as version age, have been designed to prevent unnecessary age growth in specific tasks~\cite{bastopcu2020information, yates2021Vage, buyukates2022version, delfani2024version, salimnejad2025age, champati2022detecting}. The intuition is that age need not increase when the source remains unchanged.

The aforementioned metrics, however, treat all source states equally and do not account for their \emph{contextual relevance} or goal-oriented usefulness. Information that is both fresh and accurate can still be uninformative if it does not meaningfully contribute to the system's goal. Early efforts to address this limitation include the context-aware age metric~\cite{stamatakis2019control}, which models the information source as a Markov chain with normal and alarm states. This metric penalizes the staleness of information about urgent states more heavily, highlighting that delayed updates may have unequal consequences depending on the context. Another relevant metric is the cost of actuation error (CAE)~\cite{pappas2021goal}, which captures the fact that the cost of estimation error depends not only on the physical discrepancy but also on the control cost induced by acting on the erroneous estimate. Follow-up studies~\cite{salimnejad2024real, salimnejad2023state, zakeri2025semantic} analyzed CAE for monitoring Markov sources. The work in~\cite{luo2024goal} and~\cite{luo2025semantic} extended these results to resource-constrained multi-source systems, a setting particularly relevant for multimodal systems.

An important semantic attribute motivated by many NCSs is the \emph{lasting impact} of consecutive estimation errors; that is, the longer an error persists, the more severe its consequences can become. The AoII and the cost of memory error capture this notion, though they are context-agnostic. Motivated by this perspective, the authors of~\cite{luo2024exploiting, luo2024minimizing} introduced the age of missed alarm (AoMA) and the age of false alarm (AoFA) to quantify, respectively, the persistence costs of the more urgent missed alarm error and the less important false alarm error. Moreover, the severity of an estimation error depends on both its context and duration. The significance-aware age of consecutive error (AoCE)~\cite{luo2025cost, luo2025revisiting} thus measures the \emph{urgency of lasting impact} through a set of context-aware nonlinear functions. Optimal communication policies have been derived through the lens of Markov decision theory. Prior work~\cite{luo2025cost,luo2024exploiting} has established that the optimal policy under a zero-order hold (ZOH) estimator exhibits a simple switching structure (i.e., multiple thresholds), where the transmission thresholds depend on the instantaneous estimation error. Moreover, it has been shown in~\cite{luo2026role} that, when a MAP estimator is employed, the thresholds also depend on the AoI.

\subsection{Multi-Objective MDPs}
There is a rich literature on single-objective MDPs, including Lagrangian formulations~\cite{puterman1994markov} and constrained formulations~\cite{altman1999constrained}. These formulations play a central role in communication system design~\cite{luo2025information}. In this work, we adopt a multi-objective MDP framework to characterize the value of communication. However, computing the exact Pareto front, even for two objectives, is a nontrivial task~\cite{roijers2013survey, chatterjee2006markov, chatterjee2007markov}.

\begin{table}[t!]
\caption{Notation Summary}
\label{table:notation}
\centering
\renewcommand{\arraystretch}{1.2}
\begin{tabular}{|c|l|}
\hline
\textbf{Symbol} & \multicolumn{1}{c|}{\textbf{Description}} \\
\hline
$X_{t}, \mathcal{X}, \Gamma$
& Source state, alphabet, and transition matrix \\
\hline
$I_{t}, \pi_{t}, U_{t}$
& Sensor information, decision rule, and action \\
\hline
$Y_{t}$, $\mathcal{Y}$
& Channel output and output alphabet \\
\hline
$\Theta_{t}$, $Z_{t}$ & Receiver AoI and latest received information \\
\hline
\rule{0pt}{2.3ex} $g_{t}$, $\hat{X}_{t}$ & Receiver decision rule and estimate \\
\hline
$\Delta_{t}$, $\rho_{i, j}$ & Error duration and persistence cost function \\
\hline 
$S_t$, $\mathcal{S}$ & MDP state (sufficient statistic) and state space \\
\hline 
\multirow{2}{*}{$\pi, \pi^{*}, \pi^{*}_{\beta}, \pi^{\lambda}$} & Sensor communication policy, Pareto-optimal policy, \\[-0.2em]
& constrained-optimal policy, and $\lambda$-optimal policy \\
\hline 
$c$, $\mathcal{J}(\pi)$ & Estimation cost function and its expected average \\
\hline
$f$, $F(\pi)$ & Communication cost function and its expected average \\
\hline
\rule{0pt}{2.3ex} $\ell^{\lambda}$, $\mathcal{L}^{\lambda}(\pi)$ & Lagrangian cost function and its expected average \\
\hline
$\mathcal{J}^{*}_{\beta}$, $F^{*}_{\beta}$ & {\scriptsize $\mathcal{J}^{*}_{\beta} = \mathcal{J}(\pi^{*}_{\beta})$, $F^{*}_{\beta} = F(\pi^{*}_{\beta})$} \\
\hline
$\mathcal{J}^{\lambda}, F^{\lambda}, \mathcal{L}^{\lambda}$ & {\scriptsize $\mathcal{J}^{\lambda} = \mathcal{J}(\pi^{\lambda})$, $F^{\lambda} = F(\pi^{\lambda})$, $\mathcal{L}^{\lambda} = \mathcal{L}^{\lambda}(\pi^{\lambda})$} \\
\hline
\multirow{2}{*}{$\Pi, \Pi_{\textrm{D}}, \Pi_{\beta}$} 
& Policy space, stationary deterministic policy space, \\[-0.2em]
& and feasible policy space satisfying {\scriptsize $F(\pi) \leq \beta$} \\
\hline
\multirow{2}{*}{$\mathcal{P}, \mathcal{C}, \eta$} 
& Set of Pareto-optimal policies, Pareto front, \\[-0.2em]
& and absolute slope of the Pareto front \\
\hline
\end{tabular}
\end{table}

There are two main approaches to multi-objective MDPs: single-objective methods and multi-policy methods. However, both yield only approximate or partial solutions. In the first category, multiple objectives are reduced to a single-objective problem through either a Lagrangian formulation, which combines the objectives using a weighted sum~\cite{roijers2013survey, vamplew2011empirical}, or a constrained formulation, which optimizes one objective subject to constraints on the others~\cite{altman1999constrained, miettinen1999nonlinear}. Although such methods allow standard MDP techniques to be applied, they yield only a \emph{single} Pareto point per parameter configuration, requiring multiple runs to explore different parts of the front. 

Multi-policy methods, in contrast, \emph{approximate} the Pareto front by computing a set of policies in a single run. For example, vector-valued dynamic programming~\cite{white1982multi} computes a set of nonstationary Pareto-optimal policies. This algorithm was modified in~\cite{wiering2007computing} to accelerate computation by restricting the search to stationary deterministic policies. Similarly, convex hull value iteration~\cite{barrett2008learning} computes a set of deterministic policies that form the convex hull of the Pareto front. Pareto Q-learning~\cite{van2014multi} maintains a vector Q-table to identify a set of deterministic policies. Nevertheless, existing multi-policy algorithms do not provide formal guarantees on the optimality or quality of the solutions they produce.

In this work, we characterize the structure of the Pareto front for bi-objective MDPs and show that the full front can be constructed exactly by mixing the deterministic policies at its corner points. We introduce an efficient algorithm, SPLIT, to compute the exact Pareto front.

\begin{figure}[t!]
    \centering
    \includegraphics[width=0.9\linewidth]{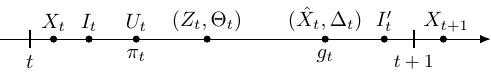}
    \caption{Time ordering of the relevant variables and decision rules.}
    \label{fig:timeline}
\end{figure}

\section{Semantics-Aware Communication Model}\label{sec:model}
This section describes the semantic communication model. The notation used in this paper is summarized in Table~\ref{table:notation}.

\subsection{System Model}\label{sec:model description}
Consider a time-slotted communication system depicted in Figure~\ref{fig:system_model}. The time ordering of the relevant variables is shown in Figure~\ref{fig:timeline}. The information source $\{X_{t}\}_{t\geq 1}$ is modeled as an irreducible Markov chain taking values in a finite alphabet $\mathcal{X}$. The chain's transition probability matrix is given by
\begin{equation*}
    \Gamma = \big[\Gamma_{i, j}\big]_{i,j \in \mathcal{X}}, \,\,\, \Gamma_{i,j} = \Pr(X_{t+1}=j {\,} | {\,} X_{t} = i). 
\end{equation*}
The probability that the source transitions from state $i$ to state $j$ in $n$ time slots is given by the $(i, j)$-th entry of $\Gamma^{n}$, i.e., 
\begin{equation*}
    \Pr(X_{t+n} = j {\,} | {\,} X_{t} = i) = \Gamma^{n}_{i,j}.
\end{equation*}
Each state of the chain represents either a quantized level of a physical process or, as is our main focus, an abstract status of a dynamical system (e.g., operating modes, component failures, or anomalies). The chain is therefore \emph{prioritized}: states differ in their relative importance to system performance.

At each time $t$, the sensor decides whether to transmit its current observation to the receiver. Let $U_{t} \in \mathcal{U} = \{0, 1\}$ denote the sensor's decision, where $U_{t} = 1$ if a transmission is attempted and $U_{t} = 0$ otherwise. We consider a practical network in which the sensor can transmit only intermittently over a packet-drop channel~\cite{schenato2008optimal, schenato2007foundations}. Specifically, each transmitted packet is either successfully delivered or lost due to channel unreliability. Let $H_{t}$ denote the channel state, where $H_{t} = 1$ represents a successful reception and $H_{t} = 0$ represents a packet drop. The sequence $\{H_{t}\}$ is modeled as an independent and identically distributed (i.i.d.) Bernoulli process with packet drop probability $\Pr(H_{t} = 0) = p_{f}$.

The output alphabet of the channel is $\mathcal{Y} = \mathcal{X} \cup \{\mathcal{E}\}$, where the symbol $\mathcal{E}$ indicates that no message was received, which occurs either when no transmission is attempted or when a transmission fails. The channel output $Y_t \in \mathcal{Y}$ is given by
\begin{equation*}
    Y_{t} = \begin{cases}
        X_{t},       & U_{t} H_{t} = 1,\\
        \mathcal{E}, & U_{t} H_{t} = 0.
    \end{cases}
\end{equation*}
The sensor is informed of the channel outcomes through an error-free feedback link. 

\subsection{Decision Rules}
The sensor has causal access to the current and past source states $X_{1:t}$, the past decisions $U_{1:t-1}$, and the delivered messages $Y_{1:t-1}$. Thus, the information available at the sensor at decision epoch $t$ (i.e., pre-communication history) is
\begin{equation*}
    I_{t} = (X_{1:t}, U_{1:t-1}, Y_{1:t-1}).
\end{equation*}
At each $t$, the sensor's decision rule is a mapping
\begin{equation*}
    \pi_{t}: \mathcal{X}^{t} \times \mathcal{U}^{t-1} \times \mathcal{Y}^{t-1} \to \mathrm{Prob}(\mathcal{U}),
\end{equation*}
where $\mathrm{Prob}(\mathcal{U})$ denotes the set of probability distributions over the action space $\mathcal{U}$. The control action is generated by
\begin{equation}
    U_{t} \sim \pi_{t}(\cdot {\,} | {\,} I_{t}), \quad t = 1, 2, \ldots \label{eq:sensor rule}
\end{equation}
Such decision rules are termed admissible and may be nonstationary, randomized, and history-dependent.

\begin{definition}
The collection $\pi = (\pi_{1}, \pi_{2}, \ldots)$ is called a \emph{communication policy}. Let $\Pi$ denote the set of admissible policies. A policy is \emph{stationary} if the decision rule is time-invariant, i.e., $\pi_{t} = \pi$ for all $t$. Moreover, a stationary policy is \emph{deterministic} if it maps the information set to a single action with probability one, i.e., $U_{t} = \pi(I_{t})$ for all $t$. Otherwise, it is called \emph{randomized}. Let $\Pi_{\textrm{D}} \subset \Pi$ denote the set of stationary deterministic policies.
\end{definition}

The receiver employs a MAP estimation rule to reconstruct the source state. This rule is assumed to be known to the sensor. Upon receiving $Y_{t}$, an estimate $\hat{X}_{t}$ is generated by the mapping $g_{t}: \mathcal{Y}^{t} \to \mathcal{X}$, defined as~\cite[Ch.~3.5]{krishnamurthy2016POMDP}
\begin{equation}
    \hat{X}_{t} = g_{t}(Y_{1:t}) = \arg\max_{x \in \mathcal{X}} B_{t}(x),\label{eq:estimator rule}
\end{equation}
where $B_{t} \in \mathrm{Prob}(\mathcal{X})$ is the receiver's posterior belief vector at time $t$, with entries
\begin{equation*}
    B_{t}(x) := \Pr(X_{t} = x {\,}| {\,} Y_{1:t}), \quad x \in \mathcal{X}.
\end{equation*}

Let $\Theta_{t}$ denote the AoI at the receiver, defined as the time elapsed since the most recent successful update, i.e.,
\begin{align}
    \Theta_{t} := t - \max \{ \tau \leq t: Y_{\tau} \neq \mathcal{E} \}.
\end{align}
In other words, $\Theta_{t}$ measures how fresh the receiver's knowledge is about the information source. By the Markov property of the source, all earlier updates are conditionally independent of $X_{t}$ given the most recent one. Thus, the receiver's posterior belief can be expressed in terms of AoI as
\begin{equation}
    B_{t}(x) = \begin{cases}
        1, &\Theta_{t} = 0, Y_{t} = x,\\
        0, &\Theta_{t} = 0, Y_{t} \neq x,\\
        \Gamma^{\Theta_{t}}_{Z_{t}, x}, &\Theta_{t} > 0, Y_{t} = \mathcal{E},
    \end{cases}
\end{equation}
where $Z_{t} = X_{t - \Theta_{t}}$ is the most recently received update, and $\Gamma^{\Theta_{t}}_{Z_{t}, x} = \Pr(X_{t} = x {\,} | {\,} X_{t-\Theta_{t}} = Z_{t})$. The MAP estimate can then be written as
\begin{equation}
    \hat{X}_{t} = g_{t}(Y_{1:t}) =  g_{t}(Z_{t}, \Theta_{t}).
\end{equation} 

\subsection{Two Competing Objectives}\label{sec:objectives}
The system must balance two competing objectives: communication efficiency and estimation cost.

\subsubsection{Communication Efficiency}
In resource-constrained systems such as sensor networks, the sensor can transmit only intermittently due to limited energy and bandwidth budgets. The communication frequency under a policy $\pi$ is defined as
\begin{equation}
    F(\pi) := \limsup_{T \to \infty}\mathbb{E}^{\pi} 
    \left[
    \frac{1}{T}\sum_{t=1}^{T} f(I_{t}, U_{t})
    \right], \label{eq:trans freq}
\end{equation}
where $f(I_{t}, U_{t}) = U_{t}$ is the per-stage communication cost.

\subsubsection{Estimation Cost}
In classical estimation systems, the system's performance is measured by average distortion, i.e.,
\begin{equation}
    \mathcal{J}_{\textrm{classic}}(\pi) := \limsup_{T \to \infty}
    \mathbb{E}^{\pi} \left[ \frac{1}{T} \sum_{t=1}^{T} d(X_{t}, \hat{X}_{t}) \right], \label{eq:classic}
\end{equation}
where $d: \mathcal{X} \times \mathcal{X} \to [0, \infty)$ is a bounded distortion measure (e.g., Hamming distortion or mean-square error) that quantifies the discrepancy between the source state and its reconstruction. The expectation in~\eqref{eq:classic} is taken with respect to a probability measure that is determined by the distributions of the source, the channel, and the communication and estimation policies.

Beyond accuracy, semantics-aware systems value the contextual relevance and goal-oriented utility of information. Let
\begin{align*}
    I^{\prime}_{t} = (X_{1:t}, U_{1:t}, Y_{1:t}, \hat{X}_{1:t})
\end{align*}
denote the complete history of system realizations up to and including time $t$ (i.e., post-communication history).

\begin{definition}
The \emph{value of information} at time $t$ is projected from the history $I^{\prime}_{t}$ by
\begin{equation*}
    \Delta_{t} = \psi_{t} (I^{\prime}_{t}) = \psi_{t}(X_{1:t}, U_{1:t}, Y_{1:t}, \hat{X}_{1:t}),
\end{equation*}
where $\psi_{t}: \mathcal{X}^{t} \times \mathcal{U}^{t} \times \mathcal{Y}^{t} \times \mathcal{X}^{t} \to \mathcal{V}$ is a projection function at time $t$, and $\mathcal{V} \subseteq [0, \infty)$ is a real-valued domain. 
\end{definition}

Then, the performance of the semantics-aware communication system can be measured by
\begin{equation}
    \mathcal{J}(\pi) := \limsup_{T \to \infty} \mathbb{E}^{\pi} 
    \left[
    \frac{1}{T} \sum_{t=1}^{T} c(X_{t}, \hat{X}_{t}, \Delta_{t})
    \right], \label{eq:average cost}
\end{equation}
where $\hat{X}_{t} = g_{t}(Z_{t}, \Theta_{t})$, and $c: \mathcal{X} \times \mathcal{X} \times \mathcal{V} \to [0, \infty)$ is a cost function, incorporating both the instantaneous error $(X_{t}, \hat{X}_{t})$ and the history-dependent value $\Delta_{t}$.

In this work, we instantiate the semantics-aware metric $c$ as the \emph{urgency of lasting impact} in consecutive errors~\cite{luo2025cost}.\footnote{We note that the analysis presented in this paper is readily applicable to most existing metrics, including those mentioned in Section~\ref{sec:related work 1}.} That is, the cost of an estimation error depends on both its contextual significance and its duration. Intuitively, the more urgent an error and the longer it persists, the more severe its consequences can become. Let $\Delta_{t} \in \mathbb{N}_{0} := \{0, 1, 2, \ldots\}$ denote the error holding time, defined recursively as\footnote{An alternative modeling choice is to let the error holding time increase continuously, regardless of changes in the error. This definition is appropriate when different errors are semantically comparable. In many applications, however, different errors may carry fundamentally distinct implications and should therefore be treated separately~\cite{costa2005mjls}; aggregating them into a single continuously increasing metric may obscure important semantic distinctions.}
\begin{equation}
    \Delta_{t} = \begin{cases}
        \Delta_{t-1} + 1, & X_{t} \neq \hat{X}_{t}, (X_{t}, \hat{X}_{t}) = (X_{t-1}, \hat{X}_{t-1}),\\
        1, & X_{t} \neq \hat{X}_{t}, (X_{t}, \hat{X}_{t}) \neq (X_{t-1}, \hat{X}_{t-1}),\\
        0, & X_{t} = \hat{X}_{t}.
    \end{cases}
\end{equation}
Then, the urgency of lasting impact is measured by
\begin{equation}
    c(X_{t}, \hat{X}_{t}, \Delta_{t}) = \rho_{X_{t}, \hat{X}_{t}}(\Delta_{t}), \label{eq:cost function}
\end{equation}
where $\rho_{i,j}: \mathbb{N}_{0} \to [0, \infty)$ are increasing functions representing the persistence costs of consecutive errors. For instance, one may impose larger base penalties (i.e., the initial severity of the error at the moment it occurs) and exponential growth functions (i.e., the urgency of the error as it persists over time) for more urgent missed alarms, and smaller base penalties and logarithmic growth functions for less important false alarms.

With a slight abuse of notation, we define the estimation cost incurred by taking action $U_{t}$ under history $I_{t}$ as
\begin{equation}
    c(I_{t}, U_{t}) = \mathbb{E} \big[
        c(X_{t}, \hat{X}_{t}, \Delta_{t}) {\,} | {\,} I_{t}, U_{t}
    \big]. \label{eq:expected estimation cost}
\end{equation}

\subsection{A Sufficient Statistic}
In optimal decision-making, it is of primary importance to identify a \emph{sufficient statistic} to reduce the policy search space. A sufficient statistic $S_{t} \subseteq I_{t}$ is the minimal information the sensor needs to retain without losing optimality. Formally, a statistic $S_{t}$ is sufficient if it satisfies
\begin{align*}
    c(I_{t}, U_{t}) = c(S_{t}, U_{t}), \,\,
    \Pr(S_{t+1} {\,} | {\,} I_{t}, U_{t}) = \Pr(S_{t+1} {\,} | {\,} S_{t}, U_{t}).
\end{align*}

\begin{lemma}
A sufficient statistic for the model described in Section~\ref{sec:model} with cost function~\eqref{eq:expected estimation cost} is
\begin{align}
    S_{t} = (X_{t}, Z_{t-1}, \Theta_{t-1}, \Delta_{t-1}).
\end{align}
\end{lemma}

As a result, there is no loss of optimality in restricting attention to \emph{Markov policies} of the form $U_{t} \sim \pi_{t} (\cdot {\,} | {\,} S_{t})$. The objectives in~\eqref{eq:trans freq} and~\eqref{eq:average cost} can then be written as
\begin{align}
    F(\pi) &= \limsup_{T \to \infty}\mathbb{E}^{\pi} 
    \left[
    \frac{1}{T} \sum_{t=1}^{T} f(S_{t}, U_{t})
    \right], \label{eq:MDP obj 2}\\
    \mathcal{J}(\pi) &= \limsup_{T \to \infty} \mathbb{E}^{\pi} 
    \left[
    \frac{1}{T} \sum_{t=1}^{T} c(S_{t}, U_{t})
    \right]. \label{eq:MDP obj 1}
\end{align}

\section{Problem Formulations}\label{sec:formulation}
\subsection{Pareto Formulation} \label{sec:pareto formulation}
The goal is to achieve an optimal tradeoff between the two competing objectives $\mathcal{J}(\pi)$ and $F(\pi)$. We formulate a bi-objective Pareto optimization problem as
\begin{equation}
    \inf_{\pi \in \Pi} \left\{ \mathcal{J}(\pi), F(\pi) \right\}. \label{problem:Pareto formulation}
\end{equation}

Problem~\eqref{problem:Pareto formulation} is a bi-objective MDP, described by the tuple $(\mathcal{S}, \mathcal{U}, P, c, f)$, where
\begin{itemize}
    \item $\mathcal{S} = \mathcal{X}^{2} \times \mathbb{N}^{2}_{0}$ denotes the space of the state process $\{S_{t}\}$. 
    \item $\mathcal{U} = \{0, 1\}$ denotes the action space.
    \item $P$ denotes the system transition kernel, where $P_{s, s^{\prime}}(u) = \Pr(s^{\prime}|s, u)$ denotes the probability of the system transitioning from state $s$ to state $s^{\prime}$ given action $u$.
    \item $c(s, u)$ and $f(s, u)$ are the per-stage cost functions.
\end{itemize}

There is generally no single ``best" policy in the Pareto sense. Instead, the aim is to characterize the Pareto front (i.e., the tradeoff curve) that contains all Pareto-optimal operating points and, more importantly, provides insights into the value of communication. Such insight is particularly useful, for example, for quantifying the marginal benefit of additional communication or for adapting to time-varying budgets.

\begin{definition}\label{def:lambda optimality}
A policy $\pi^{*} \in \Pi$ is said to be \emph{Pareto-optimal} if there exists no other policy $\pi \in \Pi$ such that
\begin{equation*}
    \mathcal{J}(\pi) \leq \mathcal{J}(\pi^{*}) \,\,\textrm{and}\,\, F(\pi) \leq F(\pi^{*})
\end{equation*}
with at least one of the above inequalities being \emph{strict}. Let $\mathcal{P} \subseteq \Pi$ denote the set of all Pareto-optimal policies. The Pareto front $\mathcal{C}$ is defined in the objective space $(F, \mathcal{J})$ as
\begin{equation*}
    \mathcal{C} = \left\{ \left( F(\pi), \mathcal{J}(\pi) \right): \pi \in \mathcal{P} \right\}.
\end{equation*}
If $\mathcal{C}$ is single-valued in $F$, we write $\mathcal{J} = \phi(F)$.
\end{definition}

\begin{definition}\label{def:value of information}
The \emph{value of communication} represents the sensitivity of the optimal system performance to changes in the communication budget, defined as the absolute slope of the Pareto front, i.e.,
\begin{equation}
    \eta(F) := \lim_{h \to 0^{+}} \frac{ \big | \phi(F + h) - \phi(F) \big|}{h}.
\end{equation}
\end{definition}

Since the Pareto front is decreasing as we will see in Section~\ref{sec:pareto}, we use the absolute value so that larger values correspond to greater marginal benefit from communication.

The following formulations serve as the basis for characterizing the Pareto front. While they have been relatively well studied in semantic communication systems, their connections to Pareto optimization remain largely unexplored.

\subsection{Constrained Formulation}\label{sec:constrained formulation}
The constrained formulation seeks to minimize the average estimation cost $\mathcal{J}(\pi)$ subject to a hard constraint on the transmission frequency $F(\pi)$. Let 
\begin{equation}
    \Pi_{\beta} := \big\{\pi \in \Pi: F(\pi) \leq \beta\big\}
\end{equation}
denote the set of feasible policies for a fixed budget $\beta$. This problem is formulated as
\begin{equation}
    \inf_{\pi \in \Pi_\beta} \mathcal{J}(\pi). \label{problem:constrained formulation}
\end{equation}

\begin{definition}\label{def:constrained-optimal}
A policy $\pi^{*}_{\beta} \in \Pi_{\beta}$ is \emph{constrained-optimal} for a given budget $\beta$ if it attains the minimum in~\eqref{problem:constrained formulation}. Let $\Pi^{*}_{\beta} \subset \Pi_{\beta}$ denote the set of constrained-optimal policies for that $\beta$. For notational simplicity, we shall write 
\begin{equation*}
    \mathcal{J}^{*}_\beta = \mathcal{J}(\pi^{*}_{\beta})\,\,\, \textrm{and}\,\,\, F^{*}_{\beta} = F(\pi^{*}_{\beta}).
\end{equation*}
\end{definition}

Problem~\eqref{problem:constrained formulation} is a Constrained MDP (CMDP). A formal approach to CMDPs is the Lagrange multiplier method, which is discussed below.

\begin{figure*}[t!]
    \centering
    \subfloat[Monotonicity and convexity (Theorem~\ref{thm:pareto-front}).]{
        \includegraphics[width=0.48\linewidth]{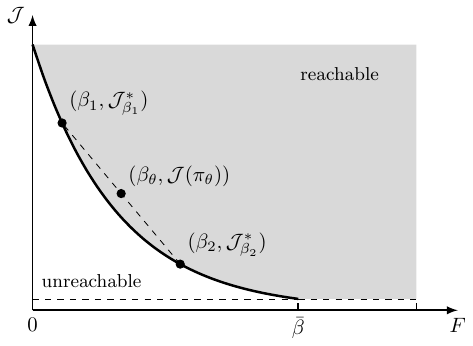}
        \label{fig:pareto front}
    }
    \hfill
    \subfloat[Piecewise linearity (Theorem~\ref{thm:pareto front lambda}).]{
        \includegraphics[width=0.48\linewidth]{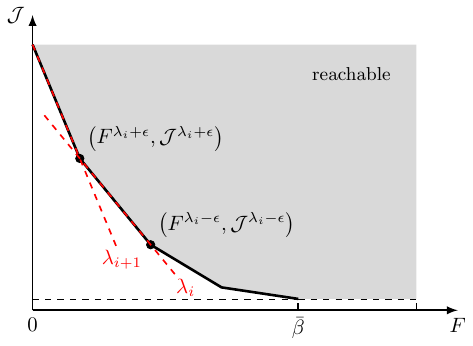}
        \label{fig:pareto front2}
    }
    \caption{Illustration of the Pareto front (solid line) obtained from (a) the constrained formulation and (b) the Lagrangian formulation. The shaded region represents the set of attainable objective values.} 
    \label{fig: pareto front total}
\end{figure*}

\subsection{Lagrangian Formulation}\label{sec:lagrangian formulation}
Let $\lambda > 0$ denote the Lagrange multiplier, which can be interpreted as the resource utilization cost associated with each transmission attempt. The Lagrangian formulation seeks to solve the following standard MDP
\begin{equation}
    \inf_{\pi \in \Pi} \mathcal{L}^{\lambda}(\pi) = \mathcal{J}(\pi) + \lambda F(\pi).\label{problem:Lagrangian formulation}
\end{equation}
This MDP can be described by the tuple $(\mathcal{S}, \mathcal{U}, P, \ell^{\lambda})$, where 
\begin{equation*}
    \ell^{\lambda}(s, u) = c(s, u) + \lambda f(s, u).
\end{equation*}

\begin{definition}\label{def:lambda-optimal}
A policy $\pi^{\lambda} \in \Pi$ is called \emph{$\lambda$-optimal} if it attains the minimum in~\eqref{problem:Lagrangian formulation} for a given $\lambda$. Let $\Pi^{\lambda} \subset \Pi$ denote the set of $\lambda$-optimal policies for that $\lambda$. We shall write 
\begin{align*}
    \mathcal{J}^{\lambda} &= \mathcal{J}(\pi^{\lambda}),~F^{\lambda} = F(\pi^{\lambda}), \\
    \mathcal{L}^{\lambda} &= \mathcal{L}^{\lambda}(\pi^{\lambda}) = \mathcal{J}^{\lambda} + \lambda F^{\lambda}.
\end{align*}
\end{definition}

We shall answer the following questions concerning the Pareto front in the next two sections.
\begin{enumerate}
    \item How are these formulations related?
    \item What is the structure of the Pareto front? 
    \item Can the value of communication be expressed explicitly?
    \item What are the characteristics of Pareto-optimal policies? More importantly, are they easy to implement?
    \item Can the front be computed exactly and efficiently?
\end{enumerate}

\section{Pareto Analysis for Bi-Objective MDPs}\label{sec:pareto}
This section presents our main results on the Pareto front for general bi-objective MDPs. We characterize the structure of the Pareto front, the value of communication, and the policies that attain points on the front. These results are then used to develop the SPLIT algorithm in Section~\ref{sec:computing}.

\subsection{Model and Assumptions}
Recall the bi-objective MDP $(\mathcal{S}, \mathcal{U}, P, c, f)$ introduced in Section~\ref{sec:pareto formulation}. We now consider a general setting with a countable state space $\mathcal{S}$ and a finite action space $\mathcal{U}$. The cost function $c(s,u) \geq 0$ represents system performance and may be unbounded, while $f(s,u) \geq 0$ is bounded and represents resource utilization. Let $F(\pi)$ and $\mathcal{J}(\pi)$ denote the two competing objectives as defined in~\eqref{eq:MDP obj 2} and~\eqref{eq:MDP obj 1}. This setting encompasses a broad class of problems including communication, control, and queueing systems~\cite{puterman1994markov, sennott1998stochastic}.

The results are primarily of interest when $\mathcal{S}$ is countably infinite and $c(s, u)$ is unbounded, but the analysis is general and also applies when $\mathcal{S}$ is finite. Our results are established for general bi-objective MDPs under the following standard assumption (see Sennott~\cite[Ch.~7]{sennott1998stochastic}).

Let $\mathcal{S}(r) = \{s \in \mathcal{S}: c(s,u) \leq r~\textrm{for some}~u \in \mathcal{U}\}$ denote the set of states whose cost can be made no larger than $r$.

\begin{assumption}\label{assump:existence}
There exists a stationary policy that induces an irreducible, positive recurrent Markov chain on $\mathcal{S}$ with finite average cost $\mathcal{J}(\pi)$. Moreover, $\mathcal{S}(r)$ is finite for every $r>0$.
\end{assumption}

The semantic communication model described in Section~\ref{sec:model} is a special case study, and a sufficient condition for Assumption~\ref{assump:existence} is given below.

\begin{lemma}\label{lem:assumption_semantic}
For the semantic communication model described in Section~\ref{sec:model}, Assumption~\ref{assump:existence} holds if, for every $i\neq j$, the persistence cost function $\rho_{i,j}$ is unbounded and satisfies 
\begin{equation}
    \lim_{\delta\to\infty} 
    \frac{\rho_{i,j}(\delta+1)}{\rho_{i,j}(\delta)}
    < \frac{1}{\Gamma_{i,i} p_f}, \quad i \in \mathcal{X},
\end{equation}
where $\Gamma_{i,i}p_f$ is the error persistence probability.
\end{lemma}
\begin{IEEEproof}
See~\cite[Thm.~1]{luo2026role}. Sufficient conditions for other metrics are discussed in~\cite[Remark~4]{luo2025cost}.
\end{IEEEproof}

\subsection{Overview of the Main Results}\label{sec:results}
The first structural result relies on the constrained formulation introduced in Section~\ref{sec:constrained formulation}.

\begin{definition}\label{def:betabar}
Let $\bar{\beta}$ be the minimum ``sufficient" budget, defined as
\begin{equation}
    \bar{\beta} := \inf \left\{ \beta \geq 0: \mathcal{J}^{*}_{\beta} = \mathcal{J}^{*}_{\infty} \right\}. \label{eq:barbeta}
\end{equation}
\end{definition}

In other words, $\bar{\beta}$ is the smallest budget beyond which additional communication provides no further improvement in estimation performance.

The next theorem establishes the \emph{monotonicity} and \emph{convexity} of the Pareto front, as illustrated in Figure~\ref{fig:pareto front}. 

\begin{theorem}\label{thm:pareto-front}
For any budget $\beta \in [0, \bar{\beta}]$, every constrained-optimal policy $\pi^{*}_{\beta}$ is Pareto-optimal. Moreover, the Pareto front is strictly decreasing and convex.
\end{theorem}
\begin{IEEEproof}
    See Section~\ref{sec:monotonicity}.
\end{IEEEproof}

\begin{remark}
Note, however, that computing the Pareto front based on Theorem~\ref{thm:pareto-front} is numerically intractable, as it requires solving infinitely many CMDPs over all possible budget values. Convexity suggests that the front can be approximated by computing the convex hull of a sufficiently dense set of Pareto-optimal points. Nevertheless, this approach is conservative, inefficient, and provides only a coarse characterization of the value of communication. 
\end{remark} 

The second structural result relies on the Lagrangian formulation introduced in Section~\ref{sec:lagrangian formulation}.

We show that $F^{\lambda}$ is a nonincreasing, piecewise-constant function of $\lambda$. Its discontinuity points identify the Lagrange multipliers at which the optimal deterministic policy changes. 

\begin{definition}\label{def:breakpoints}
Let $\Lambda = \{\lambda_{1}, \lambda_{2}, \ldots\}$, $\lambda_{1} < \lambda_{2} < \cdots$, denote the ordered set of breakpoints on $F^{\lambda}$, where each element $\lambda_{i}$ is the $i$-th breakpoint satisfying 
\begin{equation*}
    F^{\lambda_{i} - \epsilon} > F^{\lambda_{i} + \epsilon}
\end{equation*}
for any arbitrarily small $\epsilon > 0$.
\end{definition}

The next theorem shows the \emph{piecewise linearity} of the front, as illustrated in Figure~\ref{fig:pareto front2}. It also characterizes the value of communication and the policies that attain points on the front.

\begin{theorem}\label{thm:pareto front lambda}
The Pareto front has the following properties.
\begin{enumerate} 
    \item[i)] The Pareto front is piecewise linear. Its corner points are attained by deterministic $(\lambda_{i} - \epsilon)$- and $(\lambda_{i} + \epsilon)$-optimal policies for each $\lambda_{i} \in \Lambda$ and arbitrarily small $\epsilon > 0$.
    \item[ii)] The value of communication is fully characterized by $\Lambda$. In particular, for each $\lambda_{i} \in \Lambda$,
    \begin{equation}
        \eta(F) = \lambda_{i}, \,\,\, \textrm{where} \,\,\, 
        F \in \left(F^{\lambda_{i} + \epsilon}, F^{\lambda_{i} - \epsilon} \right).
    \end{equation}
    \item[iii)] Every point on the front can be realized by mixing the deterministic policies at adjacent corner points.
\end{enumerate}
\end{theorem}
\begin{IEEEproof}
    See Section~\ref{sec:linearity}.
\end{IEEEproof}

\subsection{Implications of the Main Results}
The structural properties of the Pareto front have several important implications, which we discuss below.

First, the Lagrange multiplier $\lambda$ serves not only as a measure of resource utilization and a tool for solving constrained problems, but also as a quantitative measure of the value of communication. This perspective highlights the role of Pareto optimization in semantic communications, which has not been sufficiently recognized in the literature.

Second, the randomized mixing policy (see Definition~\ref{def:randomized mixing policy}) provides a complete and efficient representation of the Pareto front: once the deterministic policies at the corner points are computed, the entire front can be constructed by mixing adjacent corner policies. Thus, by computing only the deterministic optimal policies associated with the multipliers in $\Lambda$, we recover the Pareto front exactly, rather than approximating it through a dense search over resource budgets. We leverage this result to develop an efficient algorithm in Section~\ref{sec:computing}.

Third, the solutions to the constrained formulation can also be recovered directly from the Pareto front. For any $\beta \in (F^{\lambda_{i} + \epsilon}, F^{\lambda_{i} - \epsilon})$, the constrained-optimal policy $\pi^{*}_{\beta}$ is obtained by mixing the two neighboring deterministic policies $\pi^{\lambda_{i} + \epsilon}$ and $\pi^{\lambda_{i} - \epsilon}$. Notably, this mixing policy is easy to implement: the corner policies are precomputed and stored in a lookup table, and when the operating budget or performance goal changes, the sensor simply selects the neighboring corner policies and adjusts the mixing coefficient according to~\eqref{eq:mixing coefficient}. Thus, there is no need to resolve the CMDP or store infinitely many policies. \emph{This feature is useful in resource-constrained systems such as sensor networks, where devices operate under stringent computation and memory limitations.}

Finally, the solutions to the Lagrangian formulation can be recovered directly from the front. For any $\lambda \in (\lambda_{i},\lambda_{i+1})$, the $\lambda$-optimal policy remains unchanged over the interval and coincides with $\pi^{\lambda_{i} + \epsilon}$. Thus, when the user's preference over the two competing objectives varies, the optimal operating policy can be obtained directly from the lookup table, without resolving the MDP for different multipliers.

\subsection{Proof of Theorem~\ref{thm:pareto-front}}\label{sec:monotonicity}
Recall the definitions of the constrained-optimal policy $\pi^{*}_{\beta}$ and its average costs $\mathcal{J}^{*}_{\beta}$ and $F^{*}_{\beta}$ from Definition~\ref{def:constrained-optimal}, and of the minimum sufficient budget $\bar{\beta}$ from Definition~\ref{def:betabar}.

To establish this result, we first present two key lemmas on the monotonicity and convexity of $\mathcal{J}^{*}_\beta$, as illustrated in Figure~\ref{fig:pareto front}. These properties play a central role in the subsequent analysis and algorithmic design.

\begin{lemma}\label{lemma:weakly monotonicity}
    $\mathcal{J}^{*}_{\beta}$ is weakly decreasing and convex in $\beta \geq 0$.
\end{lemma}
\begin{IEEEproof}
The proof is divided into two parts.

\textit{Part (i): Monotonicity.} Intuitively, allowing more transmissions cannot worsen the best achievable system performance. That is, the feasible set $\Pi_{\beta}$ expands as $\beta$ increases. Thus, for any $\beta_{1} < \beta_{2}$, we have $\Pi_{\beta_{1}} \subseteq \Pi_{\beta_{2}}$ and
\begin{equation}
    \mathcal{J}^{*}_{\beta_{1}} = \inf_{\pi \in \Pi_{\beta_{1}}} \mathcal{J}(\pi) \geq \inf_{\pi \in \Pi_{\beta_{2}}} \mathcal{J}(\pi) = \mathcal{J}^{*}_{\beta_{2}}.
\end{equation}
Hence, $\mathcal{J}^{*}_{\beta}$ is weakly decreasing in $\beta \in [0,\infty)$.

\textit{Part (ii): Convexity.} We need to show that straight-line interpolation lies above the Pareto front; that is, for any $\beta_{1} < \beta_{2}$ and any $\theta \in [0, 1]$, the constrained-optimal policy $\pi^{*}_{\beta_{\theta}}$ at the interpolated budget
\begin{equation*}
    \beta_{\theta} = \theta \beta_{1} + (1 - \theta) \beta_{2}
\end{equation*}
achieves a cost satisfying
\begin{equation*}
    \mathcal{J}^{*}_{\beta_{\theta}} \leq \theta \mathcal{J}^{*}_{\beta_{1}} + (1 - \theta) \mathcal{J}^{*}_{\beta_{2}}. 
\end{equation*}

As illustrated in Figure~\ref{fig:pareto front}, the key step is to construct a feasible policy $\pi_{\theta} \in \Pi_{\beta_{\theta}}$ such that 
\begin{equation}
    F(\pi_{\theta}) \leq \beta_{\theta} \,\,\, \textrm{and} \,\,\, \mathcal{J}(\pi_{\theta}) \leq \theta \mathcal{J}^{*}_{\beta_{1}} + (1 - \theta) \mathcal{J}^{*}_{\beta_{2}}. \label{eq:prove inequalities}
\end{equation}
Given such a policy, convexity follows immediately from the following chain of inequalities  
\begin{equation*}
    \mathcal{J}^{*}_{\beta_\theta} = \inf_{\pi \in \Pi_{\beta_{\theta}}} \mathcal{J}(\pi) \leq \mathcal{J}(\pi_{\theta}) \leq \theta \mathcal{J}^{*}_{\beta_{1}} + (1 - \theta) \mathcal{J}^{*}_{\beta_{2}}.
\end{equation*}

We then construct such a policy $\pi_{\theta}$ as follows. Let $\pi^{*}_{\beta_{1}}$ and $\pi^{*}_{\beta_{2}}$ denote the constrained-optimal policies corresponding to budgets $\beta_{1}$ and $\beta_{2}$, respectively.

\begin{definition}\label{def:initial mixing policy}
Define an \emph{initial mixing policy} $\pi_{\theta}$ as follows. At time $t=1$, draw a Bernoulli random variable $\Xi \sim \textsf{Bernoulli}(\theta)$. If $\Xi = 1$, execute policy $\pi^{*}_{\beta_{1}}$ for all $t \geq 1$; otherwise execute $\pi^{*}_{\beta_{2}}$ for all $t \geq 1$. 
\end{definition}

Let $J^{(i)}_{T}$ be the finite-horizon average cost under policy $\pi^{*}_{\beta_{i}}$, defined as
\begin{equation*}
    J^{(i)}_{T} = \frac{1}{T} \sum_{t=1}^{T} c(S_{t}, U_{t}), \,\,\, \textrm{where} \,\,\, U_{t} \sim \pi^{*}_{\beta_{i}, t}(\cdot {\,} | {\,} S_{t}).
\end{equation*}
Conditioned on the outcome of $\Xi$, the finite-horizon average cost under policy $\pi_{\theta}$ is given by
\begin{equation*}
    J_{T} = \begin{cases}
        J^{(1)}_{T}, &\Xi = 1,\\
        J^{(2)}_{T}, &\Xi = 0.
    \end{cases}
\end{equation*}
By construction of $\pi_{\theta}$, the random variable $J_{T}$ satisfies
\begin{equation*}
    \mathbb{E}^{\pi_{\theta}}\left[ J_{T} \right] 
    = \theta {\,} \mathbb{E}^{\pi^{*}_{\beta_{1}}}\left[J^{(1)}_{T}\right] + (1 - \theta) {\,} \mathbb{E}^{\pi^{*}_{\beta_{2}}}\left[J^{(2)}_{T}\right].
\end{equation*}
Define two sequences of real numbers $\{\bar{J}^{(i)}_{T}\}$, $i = 1, 2$, by 
\begin{equation*}
    \bar{J}^{(i)}_{T} = \mathbb{E}^{\pi^{*}_{\beta_{i}}}\left[J^{(i)}_{T}\right],
    \quad T = 1, 2, \ldots
\end{equation*}
Taking the limit superior, we obtain
\begin{align}
    \mathcal{J}(\pi_{\theta}) &= \limsup_{T \to \infty} \mathbb{E}^{\pi_{\theta}}\left[ J_{T} \right] \notag \\
    &= \limsup_{T \to \infty} \left(\theta 
    \bar{J}^{(1)}_{T} + (1 - \theta) \bar{J}^{(2)}_{T}\right) \notag \\
    &\overset{\textrm{(a)}}{\leq} \limsup_{T \to \infty} \left(\theta 
    \bar{J}^{(1)}_{T}\right) + \limsup_{T \to \infty} 
    \left((1-\theta) \bar{J}^{(2)}_{T}\right) \notag\\
    &= \theta \limsup_{T \to \infty} \bar{J}^{(1)}_{T} + (1 - \theta) \limsup_{T \to \infty} \bar{J}^{(2)}_{T} \notag\\
    &= \theta \mathcal{J}(\pi^{*}_{\beta_{1}}) + (1 - \theta) \mathcal{J}(\pi^{*}_{\beta_{2}}) \notag\\
    &\overset{\textrm{(b)}}{=} \theta \mathcal{J}^{*}_{\beta_{1}} + (1 - \theta) \mathcal{J}^{*}_{\beta_{2}}, \label{eq:prove J inequality}
\end{align}
where (a) follows from the subadditivity of the limit superior~\cite[Exercise~3.5]{rudin1976principles}, and (b) follows because $\pi^{*}_{\beta_{1}}$ and $\pi^{*}_{\beta_{2}}$ are constrained-optimal. Moreover, (a) holds with equality when both policies are stationary.

Similarly, the communication cost under $\pi_{\theta}$ satisfies
\begin{align}
    F(\pi_{\theta}) 
    &\leq \theta F(\pi^{*}_{\beta_{1}}) + (1 - \theta) F(\pi^{*}_{\beta_{2}}) \notag \\
    &=\theta F^{*}_{\beta_{1}} + (1-\theta) F^{*}_{\beta_{2}} \notag \\
    &\overset{\textrm{(c)}}{\leq} \theta \beta_{1} + (1 - \theta) \beta_{2} = \beta_{\theta}, \label{eq:prove F inequality}
\end{align}
where (c) holds because every constrained-optimal policy $\pi^{*}_\beta$ satisfies the constraint $F^{*}_{\beta} \leq \beta$. 

Combining~\eqref{eq:prove J inequality} and~\eqref{eq:prove F inequality}, we conclude that the initial mixing policy $\pi_{\theta}$ satisfies~\eqref{eq:prove inequalities}. Hence, $\pi_{\theta}$ is a valid construction, and the lemma follows.
\end{IEEEproof}

With Lemma~\ref{lemma:weakly monotonicity}, we can establish a stronger result as follows.

\begin{lemma}\label{lemma:strictly monotonicity}
    $\mathcal{J}^{*}_{\beta}$ is strictly decreasing in $\beta \in [0, \bar{\beta}]$.
\end{lemma}
\begin{IEEEproof}
We prove the result by contradiction. Recall that $\mathcal{J}^{*}_{\beta} = \mathcal{J}^{*}_{\infty}$ for all $\beta \in [\bar{\beta}, \infty)$. Suppose, to the contrary, that there exist $0 \leq \beta_{1} < \beta_{2} < \bar{\beta}$ such that 
\begin{equation*}
    \mathcal{J}^{*}_{\beta_{1}} = \mathcal{J}^{*}_{\beta_{2}}.
\end{equation*}
By Lemma~\ref{lemma:weakly monotonicity}, $\mathcal{J}^{*}_{\beta}$ is convex and weakly decreasing on $[0,\infty)$. For any $\beta_{3} \in (\beta_{2},\bar{\beta}]$, convexity gives
\begin{equation*}
    \mathcal{J}^{*}_{\beta_{2}}
    \leq \tilde{\theta} \mathcal{J}^{*}_{\beta_{1}} 
    + (1 - \tilde{\theta})
    \mathcal{J}^{*}_{\beta_{3}},
\end{equation*}
where $\tilde{\theta} = \frac{\beta_{3} - \beta_{2}}{\beta_{3} - \beta_{1}} < 1$. Since $\mathcal{J}^{*}_{\beta_{1}} = \mathcal{J}^{*}_{\beta_{2}}$, the above inequality implies $\mathcal{J}^{*}_{\beta_{2}} \leq \mathcal{J}^{*}_{\beta_{3}}$. On the other hand, weak monotonicity gives $\mathcal{J}^{*}_{\beta_{2}} \geq \mathcal{J}^{*}_{\beta_{3}}$. Hence,
\begin{equation*}
    \mathcal{J}^{*}_{\beta_{3}} = \mathcal{J}^{*}_{\beta_{2}}
    \,\,\, \textrm{for all} \,\,\, \beta_{3} \in (\beta_{2}, \bar{\beta}],
\end{equation*}
which contradicts the definition of $\bar{\beta}$ in~\eqref{eq:barbeta}. Therefore, $\mathcal{J}^{*}_{\beta}$ is strictly decreasing in $\beta \in [0, \bar{\beta}]$.
\end{IEEEproof}

\begin{figure*}[t!]
    \centering
    \subfloat[$F^\lambda$ is weakly decreasing and piecewise constant.]{
        \includegraphics[width=0.485\linewidth]{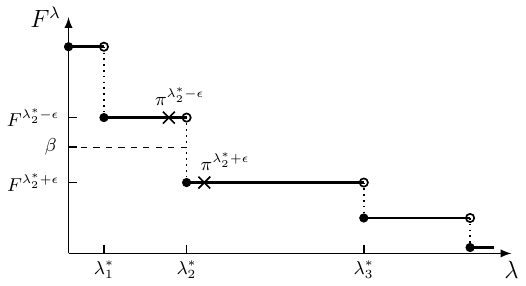}
        \label{fig:F-lambda}
    }
    \hfill
    \subfloat[$\mathcal{L}^\lambda$ is continuous, piecewise linear, and concave.]{
        \includegraphics[width=0.475\linewidth]{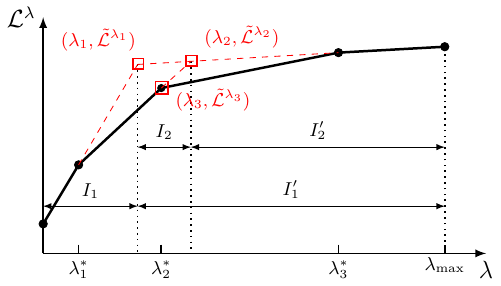}
        \label{fig:L-lambda}
    }
    \caption{Illustration of the cost functions (solid lines) in the Lagrangian formulation and the search for breakpoints.}
    \label{fig:search-algorithm}
\end{figure*}

With Lemma~\ref{lemma:strictly monotonicity}, we are now ready to establish Theorem~\ref{thm:pareto-front}.

\begin{lemma}\label{lemma:pareto-optimal-constrained}
For any $\beta \in [0, \bar{\beta}]$, $\pi^{*}_{\beta} \in \mathcal{P}$ and $\big( \beta, \mathcal{J}^{*}_\beta \big) \in \mathcal{C}$. The Pareto front is obtained as
\begin{equation}
    \mathcal{C} = \left\{ \left(\beta, \mathcal{J}^{*}_{\beta} \right): \beta \in [0, \bar{\beta} ] \right\}.
\end{equation}
The set of Pareto-optimal policies is given by
\begin{equation}
    \mathcal{P} = \left\{ \pi^{*}_{\beta} \in \Pi^{*}_{\beta}: \beta \in [0, \bar{\beta}] \right\}.
\end{equation}
\end{lemma}
\begin{IEEEproof}
We first show that $\pi^{*}_{\beta} \in \mathcal{P}$ for all $\beta \in [0, \bar{\beta}]$. Suppose that there exists $\beta \in [0, \bar{\beta}]$ such that $\pi^{*}_{\beta}$ is not Pareto-optimal. Then there exists a feasible policy $\tilde{\pi} \in \Pi_{\beta}$ such that $(F(\tilde{\pi}), \mathcal{J}(\tilde{\pi}))$ weakly dominates $(F^{*}_{\beta}, \mathcal{J}^{*}_{\beta})$, i.e.,
\begin{equation*}
    F(\tilde{\pi}) \leq F^{*}_\beta \,\, \textrm{and} \,\, 
    \mathcal{J}(\tilde{\pi}) \leq \mathcal{J}^{*}_{\beta},
\end{equation*}
with at least one strict inequality.

\textit{Case (i):} 
$F(\tilde{\pi}) \leq F^{*}_\beta$ and 
$\mathcal{J}(\tilde{\pi}) < \mathcal{J}^{*}_{\beta}$. This contradicts the fact that $\pi^{*}_{\beta}$ is constrained-optimal.

\textit{Case (ii):} 
$F(\tilde{\pi}) < F^{*}_\beta$ and 
$\mathcal{J}(\tilde{\pi}) \leq \mathcal{J}^{*}_{\beta}$. Let $\beta^{\prime} = F(\tilde{\pi})$. By optimality at $\beta^{\prime}$, we have
\begin{equation*}
    \mathcal{J}^{*}_{\beta^{\prime}} = \inf_{\pi \in \Pi_{\beta^{\prime}}} \mathcal{J}(\pi) \leq \mathcal{J}(\tilde{\pi}) \leq \mathcal{J}^{*}_{\beta}. 
\end{equation*}
By Lemma~\ref{lemma:strictly monotonicity}, 
$\mathcal{J}^{*}_{\beta}$ is strictly decreasing in $\beta \in [0, \bar{\beta}]$. Then $\mathcal{J}^{*}_{\beta^{\prime}} > \mathcal{J}^{*}_{\beta}$, which is a clear contradiction. Hence, no such $\tilde{\pi}$ exists, and $\pi^{*}_{\beta} \in \mathcal{P}$ and $(F^{*}_{\beta}, \mathcal{J}^{*}_{\beta}) \in \mathcal{C}$.

It remains to show that $F^{*}_{\beta} = \beta$ and $(\beta, \mathcal{J}^{*}_{\beta}) \in \mathcal{C}$. Suppose that $F^{*}_{\beta} < \beta$; that is, the optimal performance is achieved with a smaller budget. Let $\beta^{\prime\prime} = F^{*}_{\beta} < \beta$. Since $\pi^{*}_{\beta}$ is feasible for budget $\beta^{\prime\prime}$, we have
\begin{equation*}
    \mathcal{J}^{*}_{\beta^{\prime\prime}}
    \leq
    \mathcal{J}(\pi^{*}_{\beta})
    =
    \mathcal{J}^{*}_{\beta}.
\end{equation*}
However, $\beta^{\prime\prime} < \beta$ and Lemma~\ref{lemma:strictly monotonicity} imply
$\mathcal{J}^{*}_{\beta^{\prime\prime}} > \mathcal{J}^{*}_{\beta}$, which is a clear contradiction. Therefore, $F^{*}_{\beta} = \beta$, and hence
$(\beta, \mathcal{J}^{*}_{\beta})\in \mathcal{C}$ for every $\beta \in [0,\bar{\beta}]$. This proves the lemma.
\end{IEEEproof}

Lemma~\ref{lemma:pareto-optimal-constrained} completes the proof of Theorem~\ref{thm:pareto-front}.



\subsection{Proof of Theorem~\ref{thm:pareto front lambda}}\label{sec:linearity}
Recall the definition of the $\lambda$-optimal policy $\pi^{\lambda}$ and its average costs $\mathcal{J}^{\lambda}$ and $F^{\lambda}$ from Definition~\ref{def:lambda-optimal}.

To establish the theorem, we first present three key propositions regarding $\lambda$-optimal policies.

\begin{proposition}\label{proposition:lambda pareto optimal}
Every $\lambda$-optimal policy $\pi^{\lambda}$ is Pareto-optimal.
\end{proposition}
\begin{IEEEproof}
Suppose, to the contrary, that $\pi^{\lambda}$ is not Pareto-optimal. Then there exists a policy $\hat{\pi} \in \Pi$ such that 
\begin{equation*}
    \mathcal{J}(\hat{\pi}) \leq \mathcal{J}^{\lambda} \,\,\, \textrm{and} \,\,\, F(\hat{\pi}) \leq F^{\lambda}
\end{equation*}
with at least one strict inequality. This yields
\begin{equation*}
    \mathcal{L}(\hat{\pi}) = \mathcal{J}(\hat{\pi}) + \lambda F(\hat{\pi}) < \mathcal{L}^{\lambda},
\end{equation*}
which contradicts $\lambda$-optimality. This proves the result.
\end{IEEEproof}

The following known result, e.g., Sennott~\cite[Ch.~7]{sennott1998stochastic}, shows how to find stationary deterministic $\lambda$-optimal policies.
\begin{proposition}\label{proposition:dynamic programing}
Let Assumption~\ref{assump:existence} hold. For any $\lambda>0$, there exists a bounded function $V^{*}$ satisfying the Bellman equation
\begin{equation}
    \mathcal{L}^{\lambda} + V^{*}(s) = \min_{u \in \mathcal{U}} \left\{ \ell^{\lambda}(s, u) + \mathbb{E}\left[V^{*}(s^{\prime})| s, u \right] \right\}, \, s \in \mathcal{S}. \label{eq:bellman equation}
\end{equation}
Any deterministic policy $\pi^{\lambda} \in \Pi_{\mathrm{D}}$ attaining the minimum in~\eqref{eq:bellman equation} is $\lambda$-optimal.
\end{proposition}

The function $V^{*}$ can be computed by relative value iteration (RVI)~\cite{puterman1994markov} as follows. Let $V^{0}(s)=0$ for all $s\in\mathcal{S}$ and, for $n=1,2,\ldots$, define
\begin{subequations}
\begin{align}
    Q^{n}(s,u) &= \ell^{\lambda}(s,u) + \mathbb{E}\left[V^{n-1}(s^{\prime})|s,u \right], \label{eq:Q-factor-1}\\
    \tilde{V}^{n}(s) &= \min_{u \in \mathcal{U}}
    \{Q^{n}(s,u)\},\label{eq:Q-factor-2}\\
    V^{n}(s) &= \tilde{V}^{n}(s) - \tilde{V}^{n}(s_{\textrm{ref}}),\label{eq:Q-factor-3}
\end{align}\label{eq:Q-factor}
\end{subequations}
where $s_{\textrm{ref}} \in \mathcal{S}$ is an arbitrarily chosen reference state. Then, $\{V^{n}\}$ converges to $V^{*}$ as $n \to \infty$.

\begin{remark}
Note that it is incorrect to conclude that solving the Bellman equation~\eqref{eq:bellman equation} for all $\lambda > 0$ allows one to trace the entire Pareto front. The reasons are twofold. 
\begin{itemize}
    \item First, for a fixed $\lambda$, there may exist infinitely many $\lambda$-optimal policies, including randomized or nonstationary ones, whereas~\eqref{eq:bellman equation} returns only one deterministic policy.
    \item Second, the functions $\mathcal{J}^{\lambda}$ and $F^{\lambda}$ are discrete-valued (see Proposition~\ref{proposition:monotonicity} below). Thus, they identify only countably many isolated Pareto-optimal points.
\end{itemize}
\end{remark}

\begin{proposition}\label{proposition:monotonicity}
$F^{\lambda}$ (resp. $\mathcal{J}^{\lambda}$) is piecewise constant and weakly decreasing (resp. weakly increasing) in $\lambda$, and $\mathcal{L}^{\lambda}$ is continuous, piecewise linear, and concave (see Figure~\ref{fig:search-algorithm}).
\end{proposition}
\begin{IEEEproof}
A proof of this result can be assembled from~\cite[Lemmas~3.4--3.5]{sennott1993constrained} and~\cite[Prop.~3]{luo2025semantic}. Since our analysis and algorithm design rely on this result, we provide a consolidated proof in the Appendix for completeness.
\end{IEEEproof}

Since $F^{\lambda}$ is discontinuous, it admits a set of breakpoints. Let $\Lambda$ denote the set of ordered breakpoints (see Definition~\ref{def:breakpoints}).

We are now ready to prove Theorem~\ref{thm:pareto front lambda} using the following three lemmas, each corresponding to one of the theorem's three claims.

\begin{lemma}\label{lemma:theorem2-part1}
The Pareto front is piecewise linear. Its corner points are attained by deterministic $(\lambda_{i} - \epsilon)$- and $(\lambda_{i} + \epsilon)$-optimal policies for each $\lambda_{i} \in \Lambda$ and arbitrarily small $\epsilon > 0$.
\end{lemma}
\begin{IEEEproof}
Fix a breakpoint $\lambda_{i} \in \Lambda$. By Proposition~\ref{proposition:dynamic programing}, the policies $\pi^{\lambda_{i} - \epsilon}$ and $\pi^{\lambda_{i} + \epsilon}$ are deterministic and Pareto-optimal. Hence, the points $(F^{\lambda_{i} - \epsilon}, \mathcal{J}^{\lambda_{i} - \epsilon})$ and $(F^{\lambda_{i} + \epsilon}, \mathcal{J}^{\lambda_{i} + \epsilon})$ lie on the Pareto front. Let $\theta \in (0, 1)$ and
\begin{equation*}
    \beta_{\theta} = \theta F^{\lambda_{i} - \epsilon} + (1 - \theta) F^{\lambda_{i} + \epsilon}.
\end{equation*}
By Lemma~\ref{lemma:pareto-optimal-constrained}, we have $F^{*}_{\beta_{\theta}} = \beta_{\theta}$. 

We first show that 
\begin{equation}
   \mathcal{J}^{*}_{\beta_{\theta}} = \inf_{\pi \in \Pi_{\beta_{\theta}}} \mathcal{J}(\pi) \geq \theta \mathcal{J}^{\lambda_{i} - \epsilon} + (1 - \theta) \mathcal{J}^{\lambda_{i} + \epsilon}. \label{eq:prove assump}
\end{equation}
That is, the Pareto front lies above the line segment connecting $(F^{\lambda_{i} - \epsilon}, \mathcal{J}^{\lambda_{i} - \epsilon})$ and $(F^{\lambda_{i} + \epsilon}, \mathcal{J}^{\lambda_{i} + \epsilon})$. Suppose, to the contrary, that~\eqref{eq:prove assump} does not hold. Then
\begin{align}
    \mathcal{L}^{\lambda_{i}}(\pi^{*}_{\beta_{\theta}}) 
    &=\mathcal{J}(\pi^{*}_{\beta_{\theta}}) + \lambda_{i} F(\pi^{*}_{\beta_{\theta}}) \notag\\
    &= \mathcal{J}^{*}_{\beta_{\theta}} + \lambda_{i} \beta_{\theta} \notag\\
    &< \theta \mathcal{J}^{\lambda_{i} - \epsilon} + (1 - \theta) \mathcal{J}^{\lambda_{i} + \epsilon} \notag\\
    &\quad + \lambda_{i} \left(
    \theta F^{\lambda_{i} - \epsilon} + (1 - \theta) F^{\lambda_{i} + \epsilon}
    \right) \notag\\
    & = \theta \left( \mathcal{J}^{\lambda_{i} - \epsilon} + (\lambda_{i} - \epsilon) F^{\lambda_{i} - \epsilon} \right) \notag\\
    &\quad + (1 - \theta) \left( \mathcal{J}^{\lambda_{i} + \epsilon} + (\lambda_{i} + \epsilon) F^{\lambda_{i} + \epsilon} \right) \notag\\
    &\quad + \underbrace{\epsilon \left( \theta F^{\lambda_{i} - \epsilon} - (1 - \theta) F^{\lambda_{i} + \epsilon} \right)}_{o(\epsilon)} \notag \\
    &= \theta \mathcal{L}^{\lambda_{i} - \epsilon} + (1 - \theta) \mathcal{L}^{\lambda_{i} + \epsilon} + o(\epsilon). \label{eq:lam-inequality}
\end{align}
Taking the limit $\epsilon \to 0^{+}$ on the right-hand side of~\eqref{eq:lam-inequality} and using the continuity of $\mathcal{L}^{\lambda}$ from Proposition~\ref{proposition:monotonicity}, we obtain
\begin{align*}
    \mathcal{L}^{\lambda_{i}}(\pi^{*}_{\beta_{\theta}})  <  \mathcal{L}^{\lambda_{i}} = \inf_{\pi \in \Pi} \mathcal{L}^{\lambda_{i}}(\pi),
\end{align*}
which contradicts the optimality of $\mathcal{L}^{\lambda_{i}}$. Therefore,~\eqref{eq:prove assump} holds.

On the other hand, by the convexity of the Pareto front (see Lemma~\ref{lemma:weakly monotonicity}), we have
\begin{equation}
   \mathcal{J}^{*}_{\beta_{\theta}} \leq \theta \mathcal{J}^{\lambda_{i} - \epsilon} + (1 - \theta) \mathcal{J}^{\lambda_{i} + \epsilon}. \label{eq:prove assump 2}
\end{equation}
Combining~\eqref{eq:prove assump} and~\eqref{eq:prove assump 2} gives
\begin{equation*}
    \mathcal{J}^{*}_{\beta_{\theta}} = \theta \mathcal{J}^{\lambda_{i} - \epsilon} + (1 - \theta) \mathcal{J}^{\lambda_{i} + \epsilon}.
\end{equation*}

Thus, the point $(\beta_{\theta}, \mathcal{J}^{*}_{\beta_{\theta}})$ lies on the Pareto front. Since this holds for every $\theta \in (0, 1)$, the entire line segment also lies on the Pareto front. This proves the lemma.
\end{IEEEproof}

\begin{lemma}\label{lemma:theorem2-part2}
The value of communication is characterized by $\Lambda$. For each $\lambda_{i} \in \Lambda$ and $F \in \left(F^{\lambda_{i} + \epsilon}, F^{\lambda_{i} - \epsilon} \right)$, $\eta(F) = \lambda_{i}$.
\end{lemma}
\begin{IEEEproof}
The proof relies on the inequality that reads
\begin{align*}
    2 \epsilon F^{\lambda_{i} - \epsilon} 
    &= \mathcal{J}(\pi^{\lambda_{i} - \epsilon}) + (\lambda_{i} + \epsilon) F(\pi^{\lambda_{i} - \epsilon}) \\
    &\quad - \left( \mathcal{J}(\pi^{\lambda_{i} - \epsilon}) + (\lambda_{i} - \epsilon)F(\pi^{\lambda_{i} - \epsilon})\right) \\
    &= \mathcal{L}^{\lambda_{i} + \epsilon}(\pi^{\lambda_{i} - \epsilon}) - \mathcal{L}^{\lambda_{i} - \epsilon} \\
    &\overset{\text{(a)}}{\geq} \mathcal{L}^{\lambda_{i} + \epsilon} - \mathcal{L}^{\lambda_{i} - \epsilon} \notag\\
    &\overset{\text{(b)}}{\geq} \mathcal{L}^{\lambda_{i} + \epsilon} - \mathcal{L}^{\lambda_{i} - \epsilon}(\pi^{\lambda_{i} + \epsilon}) \\
    &= \mathcal{J}(\pi^{\lambda_{i} + \epsilon}) + (\lambda_{i} + \epsilon)F(\pi^{\lambda_{i} + \epsilon}) \\
    &\quad - \left( \mathcal{J}(\pi^{\lambda_{i} + \epsilon}) + (\lambda_{i} - \epsilon)F(\pi^{\lambda_{i} + \epsilon}) \right) \\
    & = 2\epsilon F^{\lambda_{i} + \epsilon} \geq 0,
\end{align*}
where (a) holds because the $(\lambda_{i} - \epsilon)$-optimal policy $\pi^{\lambda_{i} -\epsilon}$ is not necessarily optimal at multiplier $\lambda_{i} + \epsilon$, and (b) follows by an analogous argument. Extracting the relevant terms gives
\begin{align}
    2\epsilon F^{\lambda_{i} - \epsilon} \geq \mathcal{L}^{\lambda_{i} + \epsilon} - \mathcal{L}^{\lambda_{i} - \epsilon} \geq 2\epsilon F^{\lambda_{i} + \epsilon}\geq 0.\label{eq:prove-L-inequality}
\end{align}
Substituting
\begin{align*}
    \mathcal{L}^{\lambda_{i} + \epsilon} &= \mathcal{J}^{\lambda_{i} + \epsilon} + (\lambda_{i} + \epsilon) F^{\lambda_{i} + \epsilon},\\
    \mathcal{L}^{\lambda_{i} - \epsilon} &= \mathcal{J}^{\lambda_{i} - \epsilon} + (\lambda_{i} - \epsilon) F^{\lambda_{i} - \epsilon}
\end{align*}
into \eqref{eq:prove-L-inequality} and rearranging terms yields
\begin{align}
    (\lambda_{i} + \epsilon)\big(F^{\lambda_{i} - \epsilon}- F^{\lambda_{i} + \epsilon}\big) &\geq \mathcal{J}^{\lambda_{i} + \epsilon} - \mathcal{J}^{\lambda_{i} - \epsilon} \notag\\
    &\geq  (\lambda_{i} - \epsilon)\big(F^{\lambda_{i} - \epsilon}- F^{\lambda_{i} + \epsilon}\big).\notag
\end{align}
Since $\lambda_{i}$ is a breakpoint, we have
$F^{\lambda_{i} - \epsilon} > F^{\lambda_{i} + \epsilon}$. Dividing all terms by $F^{\lambda_{i} - \epsilon} - F^{\lambda_{i} + \epsilon}$ gives
\begin{align}
     \lambda_{i} + \epsilon \geq \frac{\mathcal{J}^{\lambda_{i} + \epsilon} - \mathcal{J}^{\lambda_{i} - \epsilon}}{F^{\lambda_{i} - \epsilon} - F^{\lambda_{i} + \epsilon}} \geq \lambda_{i} - \epsilon.
\end{align}
Taking a limit as $\epsilon \to 0^+$ yields the desired result.
\end{IEEEproof}

\begin{lemma}\label{lemma:theorem2-part3}
Every point on the Pareto front can be realized by mixing the deterministic policies at adjacent corner points.
\end{lemma}
\begin{IEEEproof}
Fix $\lambda_{i} \in \Lambda$. By Lemma~\ref{lemma:theorem2-part1}, the policies 
\begin{equation*}
    \pi^{(0)} = \pi^{\lambda_{i} + \epsilon} \,\,\, \textrm{and} \,\,\, \pi^{(1)} = \pi^{\lambda_{i} - \epsilon}
\end{equation*}
are stationary deterministic policies at adjacent corner points. Let $\theta \in (0, 1)$, and let $i_{0} \in \mathcal{S}$ be a state that is positive recurrent under both policies.

\begin{definition}\label{def:randomized mixing policy}
Define a \emph{randomized mixing policy} $\pi_{\theta}$ as follows. Each time the system visits state $i_{0}$, draw a Bernoulli random variable $\Xi \sim \mathsf{Bernoulli}(\theta)$. If $\Xi = 1$, execute $\pi^{(1)}$ until the next visit to $i_{0}$; otherwise, execute $\pi^{(0)}$ until the next visit to $i_{0}$. 
\end{definition}

Let $O_{t}$ denote the system operating mode, defined as
\begin{equation*}
    O_{t} := \begin{cases}
        0~(\textrm{\textsc{off} state}), &\textrm{if it operates under}~\pi^{(0)},\\
        1~(\textrm{\textsc{on} state}), &\textrm{if it operates under}~\pi^{(1)}.
    \end{cases}
\end{equation*}

Assume the system is initially in the \textsc{on} state and remains there for $\mathcal{T}^{(1)}_{1}$ time slots; it then jumps to the \textsc{off} state and remains \textsc{off} for $\mathcal{T}^{(0)}_{1}$ time slots. Next, it goes back to the \textsc{on} state for $\mathcal{T}^{(1)}_{2}$ time slots, and then to the \textsc{off} state for $\mathcal{T}^{(0)}_{2}$ time slots. This process repeats forever, with $\mathcal{T}^{(1)}_{n}$ being the duration of the $n$-th \textsc{on} period, and $\mathcal{T}^{(0)}_{n}$ being the duration of the $n$-th \textsc{off} period. The $n$-th \textsc{on} period followed by the $n$-th \textsc{off} period is called the $n$-th cycle. A sample path of this process is shown in Figure~\ref{fig:ARP}.

Clearly, the system regenerates after each complete cycle consisting of an \textsc{on} interval followed by an \textsc{off} interval. Let
\begin{equation*}
    \mathcal{T}_{n} = \mathcal{T}^{(1)}_{n} + \mathcal{T}^{(0)}_{n}
\end{equation*}
denote the length of the $n$-th cycle. 

\begin{figure}[t!]
    \centering
    \includegraphics[width=0.95\linewidth]{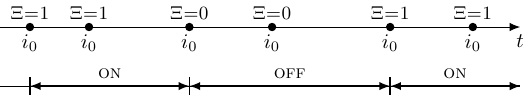}
    \caption{A sample path of the alternating renewal process.}
    \label{fig:ARP}
\end{figure}

\begin{lemma}
The sequences $\{\mathcal{T}^{(0)}_{n}\}$, $\{\mathcal{T}^{(1)}_{n}\}$, and $\{\mathcal{T}_{n}\}$ are i.i.d., and $\{O_{t}\}$ is an alternating renewal process (ARP).
\end{lemma}

Let $C^{(1)}_{n}$ and $C^{(0)}_{n}$ denote the accumulated costs during the \textsc{on} and \textsc{off} periods of the $n$-th cycle, respectively. The total accumulated cost per cycle is
\begin{equation*}
    C_{n} = C^{(1)}_{n} + C^{(0)}_{n}.
\end{equation*}
By the renewal-reward theorem~\cite[Thm.~3.6.1]{ross1996stochastic}, the long-run average cost of a renewal process is given by the ratio of the expected cost in one cycle to the expected length of that cycle. It follows that
\begin{equation*}
    \mathcal{J}(\pi^{(1)}) = \frac{\mathbb{E}[C^{(1)}_{n}]}{\mathbb{E}[\mathcal{T}^{(1)}_{n}]}, \quad 
    \mathcal{J}(\pi^{(0)}) =
    \frac{\mathbb{E}[C^{(0)}_{n}]}{\mathbb{E}[\mathcal{T}^{(0)}_{n}]}, 
\end{equation*}
and for the mixing policy $\pi_{\theta}$,
\begin{equation}
    \mathcal{J}(\pi_{\theta}) 
    = \frac{\mathbb{E}[C_{n}]}{\mathbb{E}[\mathcal{T}_{n}]} = P_{\textsc{on}} \mathcal{J}(\pi^{(1)}) + P_{\textsc{off}} \mathcal{J}(\pi^{(0)}),
\end{equation}
where $P_{\textsc{on}}$ and $P_{\textsc{off}}$ are the steady-state probabilities of being in the \textsc{on} state and the \textsc{off} state, respectively, defined as
\begin{equation*}
    P_{\textsc{on}} = \lim_{t \to \infty} \Pr(O_{t} = 1) \,\,\, \textrm{and} \,\,\, P_{\textsc{off}} =  1 - P_{\textsc{on}}.
\end{equation*}
Since $\{O_{t}\}$ is an ARP, $P_{\textsc{on}}$ satisfies~\cite[Thm.~3.4.4]{ross1996stochastic}
\begin{equation}
    P_{\textsc{on}} = \frac{\mathbb{E}[\mathcal{T}^{(1)}_{n}]}{\mathbb{E}[\mathcal{T}^{(1)}_{n}] + \mathbb{E}[\mathcal{T}^{(0)}_{n}]}. \label{eq:prob on}
\end{equation}

We next derive a closed-form expression for $P_{\textsc{on}}$. Let $N^{(k)}$ denote the number of visits to $i_{0}$ under policy $\pi^{(k)}$ before a policy switch occurs. By construction, $N^{(k)}$ follows a Geometric distribution, i.e.,
\begin{equation*}
    N^{(1)} \sim \textsf{Geometric}(1-\theta)\,\,\, \textrm{and} \,\,\, N^{(0)} \sim \textsf{Geometric}(\theta).
\end{equation*}

Let $\bar{\tau}^{(k)}$ be the expected recurrence time of state $i_{0}$ under $\pi^{(k)}$. The expected lengths of the \textsc{on} and \textsc{off} periods satisfy
\begin{align*}
    \mathbb{E}[\mathcal{T}^{(1)}_{n}] &= \mathbb{E}[N^{(1)}] \bar{\tau}^{(1)} = \frac{\bar{\tau}^{(1)}}{1 - \theta},\\
    \mathbb{E}[\mathcal{T}^{(0)}_{n}] &= \mathbb{E}[N^{(0)}] \bar{\tau}^{(0)} = \frac{\bar{\tau}^{(0)}}{\theta}.
\end{align*}
Substituting these into~\eqref{eq:prob on} yields
\begin{align}
    P_{\textsc{on}} = \frac{\theta \bar{\tau}^{(1)}}{\theta \bar{\tau}^{(1)} + (1 - \theta) \bar{\tau}^{(0)}} =: \alpha(\theta). \label{eq:mixing coefficient}
\end{align}
As a result, we have
\begin{equation}
    \mathcal{J}(\pi_{\theta})  = \alpha(\theta) \mathcal{J}(\pi^{(1)}) + \left(1 - \alpha(\theta) \right) \mathcal{J}(\pi^{(0)}).
\end{equation}

By a similar argument, $F(\pi_{\theta})$ satisfies
\begin{equation}
    F(\pi_{\theta})  = \alpha(\theta) F(\pi^{(1)}) + \left(1 - \alpha(\theta) \right) F(\pi^{(0)}).
\end{equation}
Therefore, $(F(\pi_{\theta}),\mathcal{J}(\pi_{\theta}))$ lies on the line segment connecting the two adjacent corner points.

It remains to show that every point on this segment is achievable. Since $\bar{\tau}^{(1)}$ and $\bar{\tau}^{(0)}$ are positive, the function $\alpha(\theta)$ is continuous and strictly increasing in $\theta$, with $\alpha(0) = 0$ and $\alpha(1) = 1$. Hence, as $\theta$ varies over $[0,1]$, $\alpha(\theta)$ spans the entire interval $[0,1]$. Thus, every point on the segment is realizable by some randomized mixing policy $\pi_{\theta}$.
\end{IEEEproof}

Lemma~\ref{lemma:theorem2-part3} completes the proof of Theorem~\ref{thm:pareto front lambda}.

\begin{algorithm}[t!]
\renewcommand{\algorithmicrequire}{\textbf{Input:}}
\renewcommand{\algorithmicensure}{\textbf{Output:}}
\caption{\texttt{SPLIT}$(\mathcal{B})$}
\label{alg:full}
\begin{algorithmic}[1]
\Require Operating budgets $\mathcal{B} = [F^{\lambda_{\max}}, F^{0}]$
\Ensure Pareto front $\mathcal{C}_{\mathcal{B}}$
\State $(\Lambda_{\mathcal{B}}, \Pi_{\mathcal{B}}) \gets \texttt{Insec}(\mathcal{B})$ \hfill $\triangleright$ Get breakpoints
\For{each $\lambda^{*}_{k} \in \Lambda_{\mathcal{B}}$} 
\hfill $\triangleright$ Construct front
    \State \textsc{Connect} $(F^{\lambda^{*}_{k} - \epsilon}, \mathcal{J}^{\lambda^{*}_{k} - \epsilon})$ and $(F^{\lambda^{*}_{k} + \epsilon}, \mathcal{J}^{\lambda^{*}_{k} + \epsilon})$
\EndFor
\end{algorithmic}
\end{algorithm}

\begin{algorithm}[t!]
\renewcommand{\algorithmicrequire}{\textbf{Input:}}
\renewcommand{\algorithmicensure}{\textbf{Output:}}
\caption{\texttt{Insec}$(\mathcal{B})$}
\label{alg:insec}
\begin{algorithmic}[1]
\Require $\mathcal{B} = [F^{\lambda_{\max}}, F^{0}]$
\Ensure $(\Lambda_{\mathcal{B}}, \Pi_{\mathcal{B}})$
\State \textsc{Initialize:} $I_{0} \gets [0, \lambda_{\max}]$, $\Lambda_{\mathcal{B}} \gets \emptyset$, $\Pi_{\mathcal{B}} \gets \emptyset$
\State \textsc{Breakpoints:} $(\Lambda_{\mathcal{B}}, \Pi_{\mathcal{B}}) \gets \textsc{Recurse}(I_{0})$ 
\vspace{0.3em}
\Function{Recurse}{$I_{n}$}
    \State $\lambda_{n} \gets \texttt{Update}(I_{n})$, $\pi^{\lambda_{n}} \gets \texttt{RVI}(\lambda_{n})$   \hfill $\triangleright$ Intersect
    \If{$|\mathcal{L}^{\lambda_{n}} - \tilde{\mathcal{L}}^{\lambda_{n}}| \leq \zeta$} 
    \hfill $\triangleright$ Breakpoint
        \State $\pi^{\lambda_{n} - \epsilon} \gets \texttt{RVI}(\lambda_{n} - \epsilon)$
        \State $\pi^{\lambda_{n} + \epsilon} \gets \texttt{RVI}(\lambda_{n} + \epsilon)$
        \State $\Lambda_{\mathcal{B}} \gets \Lambda_{\mathcal{B}} \cup \{\lambda_{n}\}$
        \State $\Pi_{\mathcal{B}} \gets \Pi_{\mathcal{B}} \cup \{(\pi^{\lambda_{n} - \epsilon}, \pi^{\lambda_{n} + \epsilon})\}$
    \Else \hfill $\triangleright$ Split interval
        \State $I_{n+1} \gets [\lambda_{n}^{-}, \lambda_{n}]$, $I^{\prime}_{n+1} \gets [\lambda_{n}, \lambda_{n}^{+}]$
        \State \textsc{Recurse}$(I_{n+1})$, \textsc{Recurse}$(I^{\prime}_{n+1})$
    \EndIf
\EndFunction
\end{algorithmic}
\end{algorithm}

\section{Computing the Pareto Front}\label{sec:computing}
This section leverages the structural results to compute the Pareto front for the bi-objective MDPs described in Section~\ref{sec:model}.

\subsection{Algorithm Overview}
We propose the Structured Pareto Line Intersection Tracing (\texttt{SPLIT}) algorithm, which computes the Pareto front $\mathcal{C}_{\mathcal{B}}$ over an operating budget interval $\mathcal{B} = [F^{\lambda_{\max}}, F^{0}]$. 

Let $\Lambda_{\mathcal{B}} \subseteq \Lambda$ denote the set of breakpoints of $F^{\lambda}$ within the budget interval $\mathcal{B}$, defined as
\begin{equation*}
    \Lambda_{\mathcal{B}} = \left\{ \lambda^{*}_{k} \in \Lambda: 0 \leq
    \lambda^{*}_{k} \leq \lambda_{\max} \right\}.
\end{equation*}
Let $\Pi_{\mathcal{B}}$ denote the set of optimal deterministic policy pairs corresponding to the breakpoints in $\Lambda_{\mathcal{B}}$, defined as
\begin{equation*}
    \Pi_{\mathcal{B}} = \left\{ (\pi^{\lambda^{*}_{k} - \epsilon}, \pi^{\lambda^{*}_{k} + \epsilon}): \lambda^{*}_{k} \in \Lambda_{\mathcal{B}} \right\}.
\end{equation*}

The pseudocode of \texttt{SPLIT} is summarized in Algorithm~\ref{alg:full}. It consists of two main steps.
\begin{enumerate}
    \item Identify the set of breakpoints $\Lambda_{\mathcal{B}}$ and the corresponding corner policies $\Pi_{\mathcal{B}}$ using the Intersection Search (\texttt{Insec}) method. The details of \texttt{Insec} are presented in Section~\ref{sec:insec}, and the pseudocode is given in Algorithm~\ref{alg:insec}. The key idea is to exploit the piecewise linearity and concavity of $\mathcal{L}^{\lambda}$ to identify $\Lambda_{\mathcal{B}}$ efficiently, as illustrated in Figure~\ref{fig:search-algorithm}.
    \item As illustrated in Figure~\ref{fig:pareto front2}, for each breakpoint $\lambda^{*}_{k} \in \Lambda_{\mathcal{B}}$, connect the corner points $(F^{\lambda^{*}_{k} - \epsilon}, \mathcal{J}^{\lambda^{*}_{k} - \epsilon})$ and $(F^{\lambda^{*}_{k} + \epsilon}, \mathcal{J}^{\lambda^{*}_{k} + \epsilon})$ by a line segment. Every point on this segment is attainable by the randomized mixing policy in Definition~\ref{def:randomized mixing policy}, with the mixing coefficient given in~\eqref{eq:mixing coefficient}.
\end{enumerate}

The complexity of \texttt{SPLIT} is stated below. 

Let $T_{\textrm{RVI}}$ denote the number of iterations required by RVI to compute a deterministic $\lambda$-optimal policy.
\begin{theorem}
The complexity of \texttt{SPLIT} is $\mathcal{O}(|\Lambda_{\mathcal{B}}| T_{\textrm{RVI}})$.
\end{theorem}
\begin{IEEEproof}
    See Section~\ref{sec:complexity}.
\end{IEEEproof}

\begin{remark}
The complexity of \texttt{SPLIT} is linear in the number of breakpoints. This makes the algorithm particularly efficient: although there are $|\mathcal{U}|^{|\mathcal{S}|}$ stationary deterministic policies, only a small subset appears at the corner points of the Pareto front.
\end{remark}

\begin{figure}[t!]
    \centering
    \includegraphics[width=0.6\linewidth]{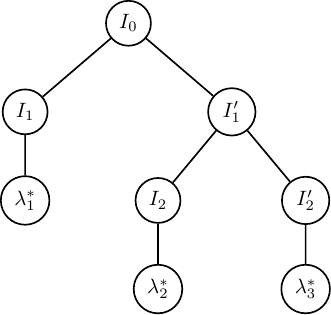}
    \caption{Binary tree representation of \texttt{SPLIT} for the case in Figure \ref{fig:search-algorithm}.}
    \label{fig:tree}
\end{figure}

\subsection{Intersection Search}\label{sec:insec}
Before introducing the \texttt{Insec} method, we first explain why directly identifying breakpoints through $F^{\lambda}$ is computationally inefficient. Recall that the mapping $F^{\lambda}$ is weakly decreasing and piecewise constant: it remains unchanged over intervals of $\lambda$ and jumps only at breakpoints, as illustrated in Figure~\ref{fig:F-lambda}. The bisection search method can be applied to locate the breakpoints. Set an initial search interval $I_{0} = [0, \lambda_{\max}]$. Bisection search divides $I_{0}$ into $I_{1} = [0, \lambda_{\max}/2]$ and $I^{\prime}_{1} = [\lambda_{\max}/2, \lambda_{\max}]$. The value of $F^{\lambda}$ is then evaluated at the endpoints of these intervals to determine whether a jump occurs inside each subinterval. This procedure is repeated for each split interval according to the following cases.
\begin{itemize}
    \item If the endpoints yield the same value of $F^{\lambda}$, then one can conclude that no breakpoint lies inside the interval, and the interval can be discarded.
    \item If the endpoints yield different values of $F^{\lambda}$, then at least one breakpoint lies inside the interval. In this case, bisection must continue subdividing the interval until its length is smaller than a prescribed tolerance.
\end{itemize}

Since $F^{\lambda}$ is flat over most of its domain and changes only at isolated breakpoints, bisection may require a large number of iterations to localize all the breakpoints accurately. The inefficiency is amplified by the fact that each evaluation of $F^{\lambda}$ requires solving an MDP at a given multiplier.

This motivates the \texttt{Insec} method~\cite{luo2025semantic}, which locates breakpoints through the Lagrangian value function $\mathcal{L}^{\lambda}$. It exploits two key structural properties: (i) $\mathcal{L}^{\lambda}$ is piecewise linear and concave, and (ii) each breakpoint of $F^{\lambda}$ corresponds to a corner point of $\mathcal{L}^{\lambda}$, as illustrated in Figure~\ref{fig:search-algorithm}. These properties allow breakpoints to be detected efficiently by tracing the intersections of supporting linear segments of $\mathcal{L}^{\lambda}$.

The algorithm proceeds as follows.
\begin{itemize}
    \item[a)] \textit{Initialization:} Choose a large multiplier $\lambda_{\max}$ and set the initial search interval $I_0 = [\lambda_0^{-}, \lambda_0^{+}] = [0, \lambda_{\max}]$.
    \item[b)] \textit{Intersection:} For a given interval $I_{n} = [\lambda_{n}^{-}, \lambda_{n}^{+}]$, construct two supporting lines of $\mathcal{L}^{\lambda}$: one passing through $(\lambda_{n}^{-}, \mathcal{L}^{\lambda_{n}^{-}})$ with slope $F^{\lambda_{n}^{-}}$, and the other passing through $(\lambda_{n}^{+}, \mathcal{L}^{\lambda_{n}^{+}})$ with slope $F^{\lambda_{n}^{+}}$. Their intersection point $(\lambda_{n}, \tilde{\mathcal{L}}^{\lambda_{n}})$ is given by
    \begin{align}
        \lambda_n &= \texttt{Update}(I_n) = \frac{\mathcal{J}^{\lambda^{+}_n} - \mathcal{J}^{\lambda^{-}_n}}{F^{\lambda^{-}_n} - F^{\lambda^{+}_n}}, \label{eq:inter-1}\\
        \tilde{\mathcal{L}}^{\lambda_{n}} &= \mathcal{J}^{\lambda^{-}_n} + \lambda_n F^{\lambda^{-}_n} \geq \mathcal{L}^{\lambda_n},\label{eq:inter-2}
    \end{align}
    as illustrated by the squares in Figure~\ref{fig:L-lambda}. Then, apply RVI to solve the Bellman equation~\eqref{eq:bellman equation} at multiplier $\lambda_{n}$ and obtain a deterministic policy $\pi^{\lambda_{n}}$.
    \item[c)] \textit{Recursion:} If the intersection point $(\lambda_n, \tilde{\mathcal{L}}^{\lambda_n})$ lies on $\mathcal{L}^\lambda$, i.e., $\tilde{\mathcal{L}}^{\lambda_n} = \mathcal{L}^{\lambda_n}$, then $\lambda_n$ is identified as a breakpoint (e.g., $(\lambda_3, \tilde{\mathcal{L}}^{\lambda_3})$ in Figure~\ref{fig:L-lambda}). In numerical implementation, this condition is replaced by the tolerance test
    \begin{equation*}
    \left|\tilde{\mathcal{L}}^{\lambda_{n}} -
    \mathcal{L}^{\lambda_{n}}\right| \leq \zeta,
    \end{equation*}
    where $\zeta > 0$ is a prescribed precision. If the test is satisfied, $\lambda_{n}$ is accepted as a breakpoint, and no further splitting of the current interval is required. Otherwise, the current interval is split into
    \begin{equation*}
        I_{n+1}=[\lambda_{n}^{-},\lambda_{n}] \,\,\, \textrm{and} \,\,\, I^{\prime}_{n+1} = [\lambda_{n}, \lambda_{n}^{+}],
    \end{equation*}
    and Step~(b) is repeated for each subinterval.
    \item[d)] \textit{Termination:} The algorithm terminates when each split interval brackets exactly one breakpoint.
\end{itemize}

\subsection{Complexity Analysis}\label{sec:complexity}
We analyze the complexity of \texttt{SPLIT} using the binary tree representation depicted in Figure~\ref{fig:tree}. In this tree,
\begin{itemize}
    \item each leaf node represents a breakpoint $\lambda^{*}_{k} \in \Lambda_{\mathcal{B}}$;
    \item each internal (non-leaf) node represents an intersection point that results in a split of the parent interval.
\end{itemize}

\texttt{Insec} maintains that each active interval brackets at least one breakpoint. Each recursive call either identifies a breakpoint or splits the current interval into two subintervals. In the corresponding binary tree, each internal node has either two children or one leaf child. Thus, the search for breakpoints is equivalent to traversing the tree until all leaves have been visited. Observe that the tree has $|\Lambda_{\mathcal{B}}|$ leaves and $2|\Lambda_{\mathcal{B}}| - 1$ internal nodes. Since the number of recursive calls equals the number of internal nodes, and each call solves the Bellman equation at a given multiplier using RVI, the complexity of \texttt{SPLIT} is $\mathcal{O}(|\Lambda_{\mathcal{B}}|T_{\textrm{RVI}})$.

In addition, the tolerance test ensures that the recursion terminates once the gap between the Lagrangian value and the intersection point becomes negligible. Therefore, the tree depth is also bounded by the discretized search space, and the worst-case complexity is $\mathcal{O}\left( (\lambda_{\max} / \zeta)  T_{\textrm{RVI}} \right)$.

\begin{figure}[t!]
    \centering
    \includegraphics[width=\linewidth]{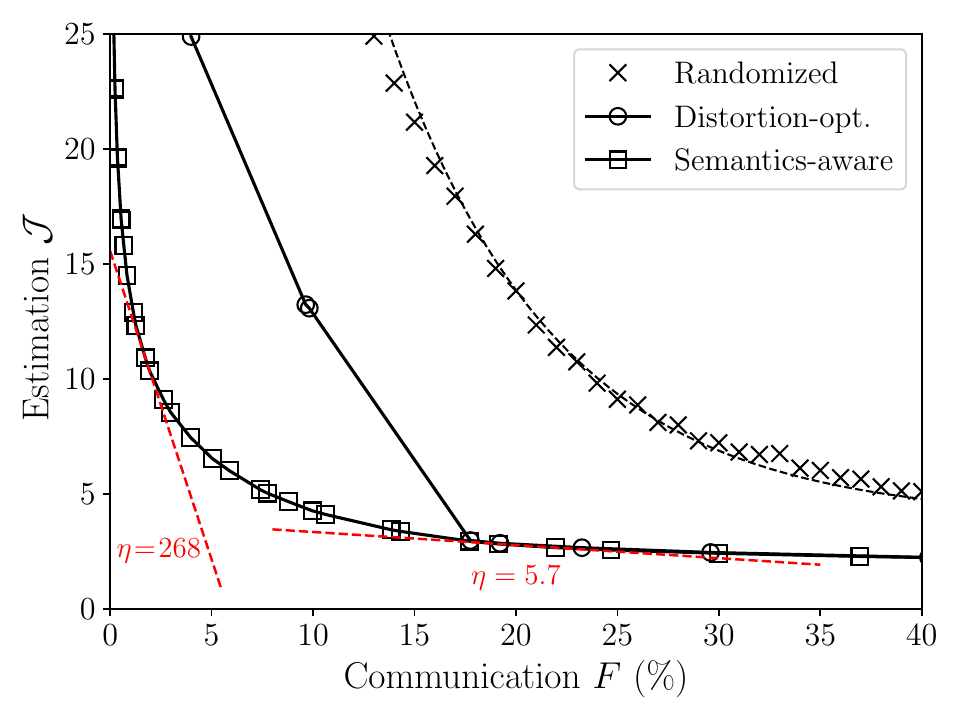}
    \caption{Estimation performance under different policies. Squares and circles represent the corner points on the Pareto fronts obtained under the semantics-aware and distortion-based policies, respectively.}
    \label{fig:front}
\end{figure}

\section{Numerical Results}\label{sec:simulations}
This section presents numerical results to corroborate the theoretical findings on the Pareto front.

We consider a dynamical system whose operating condition is abstracted as a Markov chain with a group of normal states $\{1, 2, 3, 4\}$ and a group of alarm states $\{5, 6, 7, 8\}$. The state transition probability matrix is
\begin{equation*}
\Gamma = \left[
\begin{array}{cccc|cccc}
0.65 & 0.10 & 0.00 & 0.10 & 0.05 & 0.10 & 0.00 & 0.00 \\
0.10 & 0.65 & 0.05 & 0.00 & 0.00 & 0.10 & 0.10 & 0.00 \\
0.00 & 0.05 & 0.75 & 0.10 & 0.00 & 0.00 & 0.05 & 0.05 \\
0.10 & 0.00 & 0.10 & 0.60 & 0.10 & 0.00 & 0.00 & 0.10 \\ \hline
0.10 & 0.10 & 0.00 & 0.00 & 0.70 & 0.10 & 0.00 & 0.00 \\
0.00 & 0.05 & 0.15 & 0.00 & 0.10 & 0.60 & 0.10 & 0.00 \\
0.00 & 0.00 & 0.10 & 0.10 & 0.00 & 0.10 & 0.60 & 0.10 \\
0.15 & 0.05 & 0.00 & 0.10 & 0.00 & 0.10 & 0.10 & 0.50
\end{array} \right]
\end{equation*}
The top-left, top-right, bottom-left, and bottom-right blocks represent the normal-to-normal, normal-to-alarm, alarm-to-normal, and alarm-to-alarm transitions, respectively.

The error persistence cost is modeled as
\begin{equation*}
    \rho_{i,j}(\delta) = \begin{cases}
        4.8e^{0.55\delta} + 1.2, &\textrm{missed alarm},\\
        2.4e^{0.35\delta} + 0.6, &\textrm{false alarm},\\
        1.2e^{0.15\delta} + 0.3, &\textrm{otherwise}.
    \end{cases}
\end{equation*}
The remaining parameters are set as follows: packet drop probability $p_f = 0.3$, maximum multiplier $\lambda_{\max} = 10^{5}$, and \texttt{SPLIT} search tolerance $\zeta = 10^{-3}$.

\subsection{Performance Evaluation}
For comparison, we consider the following policies.
\begin{itemize}
    \item A \textit{randomized} policy that transmits with a fixed probability in each time slot, irrespective of the system state.
    \item The \textit{distortion-optimal} policy, which minimizes the average distortion in~\eqref{eq:classic}. This policy accounts only for the instantaneous estimation error but ignores the history-dependent semantic value of information.
    \item The proposed \textit{semantics-aware} policy, which minimizes the average semantics-aware estimation cost in~\eqref{eq:average cost}.
\end{itemize}
We use \texttt{SPLIT} to compute the Pareto front for both the distortion-based and semantics-aware strategies. For the randomized policy, we take the average cost over $10^{7}$ time slots.

Figure~\ref{fig:front} compares the best achievable estimation performance under different policies. For both the distortion-based and semantics-aware strategies, the Pareto front is piecewise linear and convex, and the absolute value of its slope quantifies the value of communication. The marginal value of communication diminishes at relatively high communication frequencies, i.e., $F \geq 15\%$. In other words, communicating at a frequency higher than $F=15\%$ is inefficient, since additional communication provides little improvement in estimation performance. On the other hand, when communication is scarce, i.e., $F \leq 10\%$, careful calibration of the operating point becomes crucial. Moreover, the randomized and distortion-optimal policies exhibit significant performance degradation, particularly in the low-frequency region. 

For a given communication budget, e.g., $F=10\%$, the semantics-aware policy achieves a $70\%$ reduction in the estimation cost compared with the distortion-optimal policy. Conversely, for a target performance of $\mathcal{J}=10$, the semantics-aware policy consumes only $15\%$ of the resources used by the distortion-optimal policy and $8\%$ of those used by the randomized policy. This highlights the effectiveness of leveraging information semantics in such systems.

Consider the following design task: achieve an estimation cost no greater than $\mathcal{J} = 10$ using a communication frequency no greater than $20\%$. The constrained formulation reflects a maximalist principle: it exhausts the communication budget to minimize $\mathcal{J}$, without revealing how much communication is actually necessary. From the Pareto front, however, it is clear that the minimum communication frequency required to achieve the target performance is only $2\%$, yielding a tenfold reduction in communication.

\begin{figure*}[t!]
    \centering
    \subfloat[Relative approximation error.]{
        \includegraphics[width=0.48\linewidth]{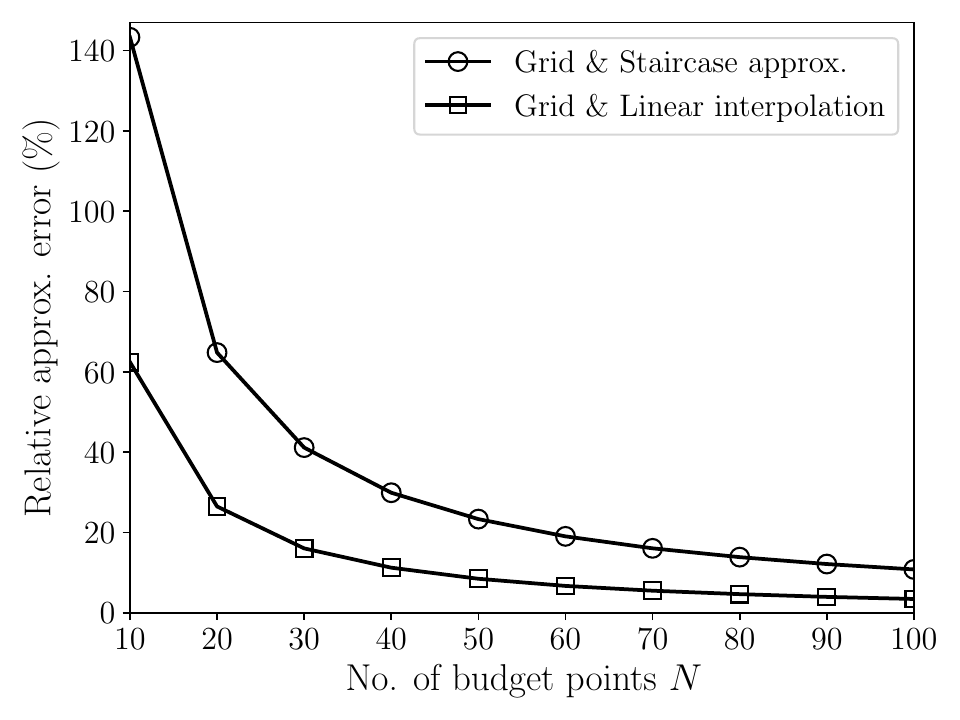}
        \label{fig:approx}
    }
    \hfill
    \subfloat[Computational complexity.]{
        \includegraphics[width=0.48\linewidth]{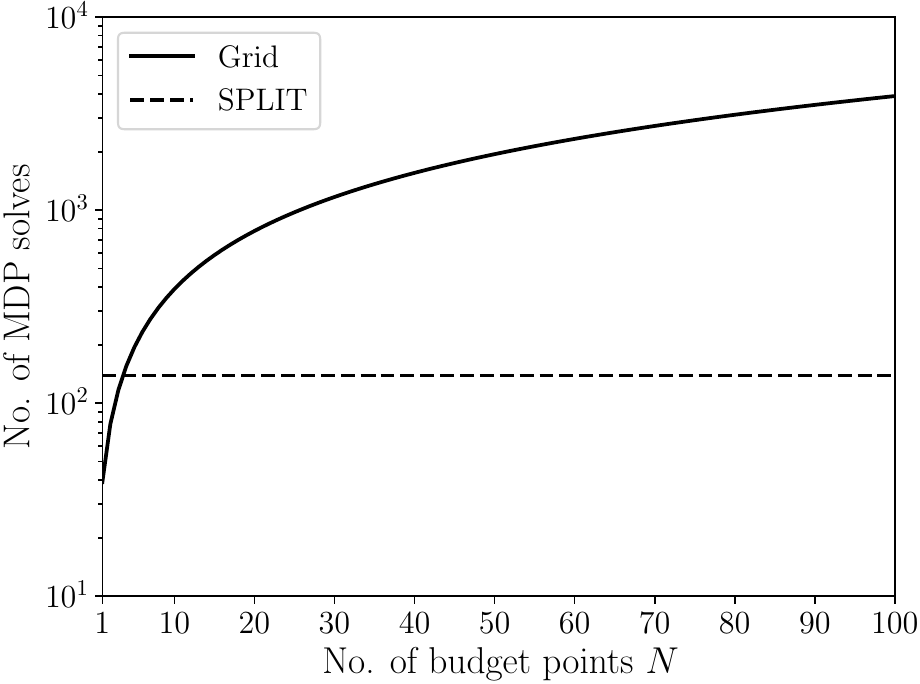}
        \label{fig:complexity}
    }
    \caption{Performance comparison between the proposed \texttt{SPLIT} and the grid-based algorithm.}
    \label{fig:comparison}
\end{figure*}

\subsection{Complexity Comparison}
As a benchmark, we consider a \emph{grid-based} algorithm, which approximates the Pareto front through a dense search over the budget interval. Specifically, the budget interval $\mathcal{B}$ is discretized into $N$ points, $\mathcal{B}_{N}= \{ \beta_{1},\beta_{2}, \ldots,\beta_{N}\}$,
and a CMDP is solved at each budget point. Each CMDP is solved via the two-layer dual problem~\cite{altman1999constrained}
\begin{equation*}
    \sup_{\lambda > 0} \inf_{\pi \in \Pi} \mathcal{L}^{\lambda}(\pi) - \lambda \beta_{n}.
\end{equation*}
The inner-layer problem is an unconstrained MDP that can be solved using RVI, while the outer-layer multiplier update is performed via bisection search. Each CMDP is solved by iterating between the two layers until the constrained-optimal policy is obtained. Further details can be found in~\cite{luo2025semantic}.

The grid-based algorithm returns only a finite set of points on the Pareto front. We consider two methods to approximate the Pareto front from these points.
\begin{itemize}
    \item \textit{Staircase approximation:} Construct a step function from these points. While conservative, this method prevents overestimation when the front shape is unknown.
    \item \textit{Linear interpolation:} Connect these points to form a piecewise linear front. This method implicitly leverages the known convexity of the Pareto front to achieve a tighter approximation.
\end{itemize}

Figure~\ref{fig:comparison} compares the performance of \texttt{SPLIT} and the grid algorithm. Figure~\ref{fig:approx} reports the relative approximation errors of the above approximation methods. The results show that the grid algorithm requires a sufficiently dense search to achieve a close approximation of the Pareto front. Figure~\ref{fig:complexity} compares the complexity of the algorithms in terms of the number of MDP solves, i.e., RVI calls. The grid algorithm uses thousands of MDP solves to reach the desired accuracy. \emph{Notably, \texttt{SPLIT} computes the exact Pareto front with the same order of complexity as solving a single CMDP.}

Table~\ref{table:complexity} reports the number of MDP solves versus the number of source states $|\mathcal{X}|$. The grid algorithm requires $3900$ MDP solves for all tested values of $|\mathcal{X}|$ with a fixed number of budget points $N = 100$. The number of MDP solves is fixed across the tests because it depends only on the number of budget points and the bisection resolution. However, a finer grid is required to maintain accuracy as $|\mathcal{X}|$ increases. In contrast, \emph{\texttt{SPLIT} requires more than an order of magnitude fewer MDP solves across all cases.} This efficiency gain is particularly significant when the problem size is large and each MDP solve requires substantial computation time. These results demonstrate the efficiency and scalability of \texttt{SPLIT}.

\begin{table}[t]
\centering
\caption{Number of MDP solves versus the number of source states.}
\label{table:complexity}
\begin{tabular}{ccccccc}
\hline
\multirow{2}{*}{Algorithm} & \multicolumn{6}{c}{Number of source states $|\mathcal{X}|$} \\
\cline{2-7}
\rule{0pt}{2.ex} & $2$ & $4$ & $6$ & $8$ & $10$ & $12$ \\
\hline
\rule{0pt}{2.ex} Grid ($N \hspace{-0.2em}=\hspace{-0.2em} 100$) & 3900 & 3900 & 3900 & 3900 & 3900 & 3900 \\
\texttt{SPLIT} & 58   & 73   & 139  & 146  & 208  & 257  \\
\hline
\end{tabular}
\end{table}

\section{Concluding Remarks}
\label{sec:conclusion}
This paper investigates the value of communication in goal-oriented semantic communications through the canonical problem of remote state estimation of Markov sources. The optimal communication design is formulated as a bi-objective MDP, and the value of communication is defined as the absolute slope of the resulting Pareto front.

Our analysis reveals that the Pareto front for bi-objective MDPs is strictly decreasing, convex, and piecewise linear. Each corner point corresponds to a stationary deterministic policy, and the entire front can be constructed by randomly mixing the deterministic policies at neighboring corner points. The value of communication is governed by a set of Lagrange multipliers. Moreover, the solutions to the Lagrangian and constrained formulations can be recovered directly from the front. This feature enables online adaptation when the system budget or preference goal changes and is useful in resource-constrained systems where devices operate under stringent computation and memory limitations.

Building on these structural insights, we develop SPLIT, a fast and provably optimal algorithm that computes the exact Pareto front with complexity linear in the number of corner points. Numerical results demonstrate that SPLIT achieves the same order of complexity as solving a single constrained problem, while existing approximation methods require more than an order of magnitude more MDP solves. Furthermore, the minimum communication required to achieve a target performance level is significantly lower than what a constrained formulation would consume. This highlights the effectiveness of Pareto-optimal semantic communication design.

\renewcommand{\theequation}{A\arabic{equation}}
\setcounter{equation}{0} 
\appendix
Proposition~\ref{proposition:monotonicity} is established using the following lemmas.

\begin{lemma}[Monotonicity]
$\mathcal{L}^{\lambda}$ and $\mathcal{J}^{\lambda}$ are weakly increasing in $\lambda$, while $F^{\lambda}$ is weakly decreasing in $\lambda$.
\end{lemma}
\begin{IEEEproof}
For any $\lambda > 0$ and $\epsilon > 0$, let $\pi^{\lambda}$ and $\pi^{\lambda + \epsilon}$ denote the $\lambda$-optimal and $(\lambda + \epsilon)$-optimal policies. Then, we have
\begin{align}
    \mathcal{L}^{\lambda + \epsilon} - \mathcal{L}^{\lambda} 
    &= \mathcal{L}^{\lambda + \epsilon}(\pi^{\lambda + \epsilon}) - \mathcal{L}^{\lambda}(\pi^{\lambda}) \notag\\
    &\geq \mathcal{L}^{\lambda + \epsilon}(\pi^{\lambda + \epsilon}) - \mathcal{L}^\lambda(\pi^{\lambda + \epsilon}) \notag\\
    & = \mathcal{J}(\pi^{\lambda + \epsilon}) + (\lambda + \epsilon) F(\pi^{\lambda + \epsilon}) \notag\\
    &\quad - \left( \mathcal{J}(\pi^{\lambda + \epsilon}) + \lambda F(\pi^{\lambda + \epsilon}) \right) \notag\\
    & = \epsilon F^{\lambda + \epsilon} \geq 0.\label{eq:appendix-a}
\end{align}
Hence, $\mathcal{L}^{\lambda}$ is weakly increasing in $\lambda$. Similarly, we have
\begin{align}
    \mathcal{L}^{\lambda + \epsilon} - \mathcal{L}^{\lambda} 
    &\leq \mathcal{L}^{\lambda + \epsilon}(\pi^{\lambda}) - \mathcal{L}^{\lambda}(\pi^{\lambda}) \notag\\
    & = \mathcal{J}(\pi^{\lambda}) + (\lambda + 
    \epsilon) F(\pi^{\lambda}) \notag\\
    &\quad - \left(\mathcal{J}(\pi^{\lambda}) + \lambda F(\pi^{\lambda}) \right) \notag\\
    & = \epsilon F^{\lambda}.\label{eq:appendix-b}
\end{align}
Combining~\eqref{eq:appendix-a} and~\eqref{eq:appendix-b}, we obtain
\begin{equation}
    0 \leq \epsilon F^{\lambda + \epsilon} \leq \mathcal{L}^{\lambda + \epsilon} - \mathcal{L}^{\lambda} \leq \epsilon F^{\lambda}, \label{eq:appendix-c}
\end{equation}
which implies that $F^{\lambda}$ is weakly decreasing in $\lambda$.

It remains to show that $\mathcal{J}^{\lambda}$ is weakly increasing. Suppose, to the contrary, that $\mathcal{J}^{\lambda}$ is decreasing in $\lambda$. Then we must have
\begin{equation}
    \mathcal{L}^{\lambda} = \mathcal{J}^{\lambda} + \lambda F^{\lambda} > \mathcal{J}^{\lambda + \epsilon} + \lambda F^{\lambda + \epsilon} = \mathcal{L}^\lambda(\pi^{\lambda + \epsilon}),
\end{equation}
which contradicts the $\lambda$-optimality of $\pi^{\lambda}$. Therefore, $\mathcal{J}^{\lambda}$ is weakly increasing in $\lambda$.
\end{IEEEproof}

\begin{lemma}[Continuity and concavity]
$\mathcal{L}^{\lambda}$ is concave and continuous on $(0, \infty)$.
\end{lemma}
\begin{IEEEproof}
For any $0 < \lambda_{1} < \lambda_{2}$ and $0 \leq \theta \leq 1$, let $\lambda_{\theta} = \theta \lambda_{1} + (1 - \theta) \lambda_{2}$. Then we have 
\begin{align}
    \mathcal{L}^{\lambda_{\theta}} 
    &= \mathcal{J}^{\lambda_{\theta}} + \lambda_{\theta} F^{\lambda_{\theta}}\notag\\
    &= \mathcal{J}^{\lambda_{\theta}} + \left(\theta \lambda_{1} + (1 - \theta) \lambda_{2} \right) F^{\lambda_{\theta}} \notag\\
    &= \theta \left(\mathcal{J}^{\lambda_{\theta}} + \lambda_{1} F^{\lambda_{\theta}} \right) + (1 -\theta) \left(\mathcal{J}^{\lambda_{\theta}} + \lambda_{2} F^{\lambda_{\theta}}\right) \notag\\
    &\geq \theta \mathcal{L}^{\lambda_{1}} + (1 - \theta) \mathcal{L}^{\lambda_{2}}.
\end{align}
This establishes the concavity of $\mathcal{L}^{\lambda}$ on $(0, \infty)$. 

Let $F_{\max} = \max_{\lambda > 0} F^{\lambda} < \infty$. It follows from~\eqref{eq:appendix-c} that for any $0 < \lambda_{1}, \lambda_{2} < \infty$,
\begin{align}
    \big| \mathcal{L}^{\lambda_{1}} - \mathcal{L}^{\lambda_{2}} \big| 
    &\leq \max \{F^{\lambda_{1}}, F^{\lambda_{2}} \} \big| \lambda_{1} - \lambda_{2} \big| \notag\\
    &\leq F_{\max} \big| \lambda_{1} - \lambda_{2} \big|.
\end{align}
Hence $\mathcal{L}^{\lambda}$ is Lipschitz and therefore continuous on $(0, \infty)$.
\end{IEEEproof}

\begin{lemma}[Piecewise linearity]
$\mathcal{L}^{\lambda}$ is piecewise linear, and $\mathcal{J}^{\lambda}$ and $F^{\lambda}$ are piecewise constant.
\end{lemma}
\begin{IEEEproof}
We prove this by induction on the value iteration recursion in~\eqref{eq:Q-factor}. Initialize 
\begin{equation*}
    Q^{0}(s, u) = \ell^{\lambda}(s,u) = c(s, u) + \lambda u.
\end{equation*}
Clearly, $Q^{0}(s,u)$ is linear in $\lambda$. Since the pointwise minimum of linear functions is piecewise linear, both $\tilde{V}^{0}(s) = \min_{u} \{Q^{0}(s, u)\}$ and $V^{0}(s) = \tilde{V}^{0}(s) - \tilde{V}^{0}(s_\textrm{ref})$ are piecewise linear in $\lambda$ for all $s$ and $u$.

Assume as the induction hypothesis that $Q^{n}(s, u)$ is piecewise linear for all $s\in\mathcal{S}$ and $u\in\mathcal{U}$. Since weighted sums of piecewise linear functions are piecewise linear, it follows that
\begin{equation*}
    Q^{n+1}(s, u) = \ell^{\lambda}(s,u) + \mathbb{E} \left[ V^{n}(s^{\prime}) | s, u \right]
\end{equation*}
is piecewise linear in $\lambda$ for all $s$ and $u$. Since this property is preserved as $n \to \infty$, the optimal value functions $V^{*}(s)$ and $Q^{*}(s, u)$ are piecewise linear. Consequently, 
\begin{equation*}
    \mathcal{L}^{\lambda} = \min_{u} Q^{*}(s, u) - V^{*}(s)
\end{equation*}
is piecewise linear in $\lambda$. 

From~\eqref{eq:appendix-c}, the right-hand derivative of $\mathcal{L}^{\lambda}$ satisfies
\begin{equation}
    \lim_{\epsilon \to 0^{+}}F^{\lambda + \epsilon} \leq \lim_{\epsilon \to 0^{+}} \frac{\mathcal{L}^{\lambda + \epsilon} - \mathcal{L}^{\lambda}}{\epsilon} \leq F^{\lambda}.
\end{equation}
Thus, at every non-corner point, the derivative of $\mathcal{L}^{\lambda}$ exists and equals $F^{\lambda}$. Since $\mathcal{L}^{\lambda}$ is piecewise linear, its derivative is constant on each linear segment. Therefore, $F^{\lambda}$ is piecewise constant, and so is $\mathcal{J}^{\lambda}$.
\end{IEEEproof}


\bibliographystyle{IEEEtran}
\bibliography{ref}

\end{document}